\documentclass[twocolumn,english,preprintnumbers,showkeys,showpacs,nofootinbib,prl]{revtex4}

\usepackage{cancel}
\usepackage{epsfig}
\usepackage{mathrsfs}
\usepackage{graphicx}
\usepackage{dcolumn}
\usepackage{bm}
\usepackage{hyperref}
\usepackage{backref}
\usepackage{verbatim}
\usepackage[all, knot]{xy}
\xyoption{arc}
\usepackage{amsmath}
\usepackage{amssymb}
\usepackage{revsymb}
\usepackage{slashed}
\usepackage[usenames]{color}
\usepackage{tikz}

\begin{document}

\title{Quantum lattice gas model of spin-2 Bose-Einstein condensates \\
and closed-form  analytical continuation of nonlinear interactions in spin-2 superfluids}

\author{Jeffrey Yepez}\email{yepez@hawaii.edu}
\date{September 7, 2016, Revised August 7, 2017}
 \affiliation{
  Department\;of\;Physics\;and\;Astronomy,  University\;of\;Hawai`i\;at\;M\=anoa, Watanabe\;Hall,\;2505\;Correa\;Road, Honolulu,\;Hawai`i\;96822
}

\begin{abstract}
Presented  is an unitary operator splitting method for handling the spin-density interaction in spinor Bose-Einstein condensates. The zero temperature behavior of  a spinor BEC is given by mean field theory, where the  Hamiltonian includes a nonlinear hyperfine spin interaction. This hyperfine interaction has a diagonal probability-density term (leading to the usual Gross-Pitaevskii type equation of motion) but also has a nondiagonal spin-density term.  Since the  $F=2$ spinor BEC  (spin-2 BEC) has  a non-Abelian superfluid phase (nonperturbative cyclic phase  in the strong spin-density coupling regime), an infinite-order expansion of the quantum evolution operator is needed for quantum simulation applications.  An infinite-order expansion,  obtained by analytical continuation and expressed in analytically closed form,  for the spin-2 BEC is presented.
\end{abstract}

\pacs{67.85.Fg,03.75.Mn,03.67.Ac,03.65.Aa}

\keywords{spinor Bose-Einstein condensates, non-Abelian gauge group, strongly-coupled ultracold quantum gas, multicomponent Gross-Pitaevskii equations, quantum computing, quantum simulation}
\maketitle

\section{Introduction}

Through  magneto-optical trapping combined with laser and evaporative cooling \cite{RevModPhys.74.1131}, ultracold quantum gases can be cooled within the timespan of a couple seconds  to the submicrokelvin temperature range for phase change to a Bose-Einstein condensate (BEC), theoretically predicted 90 years ago \cite{1925SPAW1_3}. Two decades ago,  BECs were experimentally realized using dilute atomic vapors \cite{Anderson:1995p2538,PhysRevLett.75.3969,PhysRevLett.77.416}.  Since their initial  realization using magneto-optical trapping, new trapping techniques have emerged, for example using an atom chip in an encapsulated vacuum cell \cite{PhysRevA.70.053606} 
 and optical lattices \cite{PhysRevLett.91.250402,PhysRevLett.92.130403,PhysRevLett.111.185302,PhysRevLett.107.255301,ncomms3615}.
Improvements in lasers and atom-chip vacuum cells has allowed for an   experimental  test system to  fit into a small platform \cite{PhysRevA.87.053417} not much bigger than the size of computer workstations when they were originally introduced.  

Spinor  BECs can be  reliably reproduced at every few seconds,  with repeated  observations made through high numerical aperture contrast imaging.   Such experimental systems are useful for studying superfluidity and topological solitons in spinor BECs. A spinor BEC has a $(2f+1)$-multiplet quantum matter field that represents its spin degrees of freedom in the Zeeman manifold. An example spin-2 BEC is an ultracold quantum gas of alkali Rubidium-87 atoms with total angular momentum $F=L+S+I=2\hbar$, when the orbital angular momentum  is $L=0$, since the intrinsic electron spin is $S=\hbar/2$  and the nuclear spin of ${}^{87}$Rb is $I={3\hbar}/{2}$.  The Zeeman  manifold of hyperfine level is $2F+1=5$ dimensional.     A spin-2 BEC has ferrromagnetic, polar, and cyclic phases.   The cyclic phase  of a spin-2 BEC  is analogous to $d$-wave Bardeen-Cooper-Schrieffer (BCS) superfluid \cite{PhysRevA.61.033607}, and it is this phase that admits non-Abelian quantum vortex solitons. So when confined in a  magneto-optical trap, an atom-chip trap,  or an optical lattice,  a spin-2 BEC can be used to study  non-Abelian superfluid dynamics.    Experimental realizations of a spin-2 BEC provides a way to study a non-Abelian gauge theory in a table-top platform.    Furthermore, designing the next generation of  small-platform BEC test systems for   applications such as quantum computing \cite{PhysRevA.86.032323}, analog quantum simulation \cite{PhysRevLett.85.4643,PhysRevLett.107.275301,PhysRevLett.107.255301,PhysRevLett.105.190403,Tagliacozzo2013160}, studying topological solitons in spinor BECs \cite{PhysRevA.64.043612},  and BEC interferometry \cite{PhysRevLett.94.090405,PhysRevA.72.021604,PhysRevLett.98.030407,PhysRevLett.98.180401,PhysRevA.75.063603,PhysRevA.77.043604,PhysRevA.86.053401}   
 all require accurate time-dependent quantum simulation of spinor BECs to advance the state of the art. 
 
 The purpose of this communication is to present a quantum computing method to accurately model a spin-2 BEC. The method  is based on an  infinite-order expansion of the hyperfine spin-density interaction.  It is useful for modeling the time-dependent dynamics and interaction of solitons in the non-Abelian superfluid phase of a spin-2 BEC.  Infinite-order expansions  of the spin density and singlet-pair density (nondiagonal) parts of the hyperfine interaction  unitary operator, denoted   $U^\text{\tiny nd}_f$,  are presented in analytically closed form for  the spin $f=2$ BEC in the strongly-coupled  nonperturbative regime.  An exact expansion of $U^\text{\tiny nd}_{f=1}$ is provided as a warm-up exercise before treating the spin-2 case. 
 
 The method is a quantum lattice gas model that uses an operator splitting technique that avoids the Baker-Campbell-Hausdorff (BCH) catastrophe that normally occurs because the kinetic energy operator in the free part of the Hamiltonian does not commute with the potential energy  (nonlinear point-contact) operator in the interaction part of the Hamiltonian for a spinor BEC.  So the operator splitting method in the quantum lattice gas algorithm does not require the   Lie-Trotter product formula \cite{Trotter_JSTOR_1959} to separate the kinetic energy  and interaction potential energy operators during the time evolution.   Instead, each component of the spin-$f$ bosonic field is represented by a pair of spin-1/2 fermionic fields. 
 The method is useful for computational physics applications of spin-2 BECs on parallel computers.  The numerical results obtained using this quantum lattice gas algorithm will be reported in an accompanying manuscript on interacting non-Abelian quantum vortices.

\section{Spinor BECs}

The effective Hamiltonian of a spinor BEC constructed from a bosonic alkali atom of mass $m$ follows from mean-field theory \cite{ueda_10}.  Here a quantum computing algorithm is presented for the many-body system. The main approximation is reducing the boson-boson interaction to point-contact form. Each hyperfine spin state can be treated as a separate bosonic species. For integer atomic spin $f$, there are $2f+1$  hyperfine states labeled with quantum numbers $m=f, f-1, \dots, 0, \dots, 1-f, -f$. Thus, a hyperfine multiplet spinor bosonic field operator, say $\hat\varphi$, has $2f+1$ operator components
\begin{equation}
{\hat \varphi}= 
\begin{pmatrix}
      {\hat \varphi}_f    &
      {\hat \varphi}_{f-1} &
      \cdots &
      {\hat \varphi}_{1-f} &
      {\hat \varphi}_{-f}  
\end{pmatrix}^\text{T}.
\end{equation}
Neglecting an external trapping potential and a background uniform magnetic field, 
the diagonal part of the Hamiltonian  is 
\begin{equation}
\label{diagonal_energy_functional}
H_\text{diag}[{\hat \varphi}]= \int d^3 r \sum_{m} {\hat \varphi}_m^\dagger
 \left[
 -\frac{\hbar^2}{2m}\nabla^2 - \mu_m
 +
  \frac{g_0}{2} ({\hat \varphi}^\dagger {\hat \varphi})
 \right]
 {\hat \varphi}_{m},
\end{equation}
for $m=-f, \dots, f$, and (up to $f=2$) the  nondiagonal part is
\begin{equation}
\label{nondiagonal_energy_functional}
H_\text{nondiag}[{\hat \varphi}]=  \frac{g_1}{2\hbar^2}\int d^3 r
 {\hat{\bm{F}}}^2
+
 \frac{g_2}{2}\int d^3 r
  \hat A_{00}^\dagger(\bm{r})   \hat A_{00}(\bm{r})
,
\end{equation}
where the spin density vector is
\begin{equation}
\label{spin_density_vector}
{\hat{\bm{F}}}(\bm{r}) =  {\hat \varphi}^\dagger(\bm{r}) \bm{f} {\hat \varphi}(\bm{r}) = \sum_{mm'}{\hat \varphi}_m^\dagger(\bm{r}) \bm{f}_{mm'} {\hat \varphi}_{m'}(\bm{r}),
\end{equation}
 where the spin vector is $\bm{f} = (f_x, f_y, f_z)$,
  and where the singlet-pair density  is $\hat N^{00}(\bm{r})=  \hat A_{00}^\dagger(\bm{r})   \hat A_{00}(\bm{r})$.  The annihilation operator for the singlet-pair is
\begin{equation}
\label{number_density}
 \hat A_{00}(\bm{r})
 =
  {\hat \varphi}^\dagger(\bm{r}) N_{00} {\hat \varphi}(\bm{r}) ,
\end{equation}
where the components of $N_{00}$ are
\begin{equation}
\label{N00_matrix}
N^{00}_{mm'}
=
\frac{(-1)^{f-m}}{\sqrt{2f+1}}\delta_{m,-m'}.
\end{equation}
The equal-time commutators are
\begin{align}
\begin{split}
 [\hat\varphi_m(\bm{x}_a), \hat\varphi_n^\dagger(\bm{x}_b)]
 &=\delta^{(3)}(\bm{x}_a-\bm{x}_b))\delta_{mn}
 \\
[ \hat\varphi_m(\bm{x}_a), \hat\varphi_n(\bm{x}_b)]
 &=0
 \qquad\quad
[ \hat\varphi_m^\dagger(\bm{x}_a), \hat\varphi_n^\dagger(\bm{x}_b)]=0
 ,
 \end{split}
  \end{align}
   where $m$ and $n$ denote the Zeeman levels.  %

    The  full Hamiltonian  is  $H[{\hat \varphi}]=H_\text{diag}[{\hat \varphi}]+H_\text{nondiag}[{\hat \varphi}]$, and the equation of motion for the $m$th component is
\begin{equation}
\label{formal_equation_of_motion}
i \hbar \partial_t {\hat \varphi}_m
  =  
 \frac{\delta H_\text{diag}[{\hat \varphi}]}{\delta {\hat \varphi}_m^\dagger}
+
 \frac{\delta H_\text{nondiag}[{\hat \varphi}]}{\delta {\hat \varphi}_m^\dagger}.
\end{equation}
Quantum simulations using  $2f+1$ coupled Gross-Pitaevskii equations  
 \cite{PhysRevA.83.023619} need a tractable way to handle  
the nondiagonal part of the hyperfine spin-density interaction in (\ref{formal_equation_of_motion}), 
which is
\begin{equation}
\label{nondiagonal_equation_of_motion}
 \frac{\delta H_\text{nondiag}[{\hat \varphi}]}{\delta {\hat \varphi}_m^\dagger} =\frac{ g_1}{\hbar^2} {\hat{\bm{F}}} \cdot \frac{\delta {\hat{\bm{F}}}}{\delta {\hat \varphi}_m^\dagger}
 +
 \frac{ g_2}{2}  \frac{\delta  \hat A_{00}^\dagger  }{\delta {\hat \varphi}_m^\dagger} \hat A_{00}
.
\end{equation}
Since
 $\frac{ g_1}{\hbar^2}   {\hat{\bm{F}}}. \left(\frac{\delta ({\hat \varphi}^\dagger \bm{f} {\hat \varphi})}{\delta {\hat \varphi}^\dagger}\right)=\frac{ g_1}{\hbar^2}  {\hat{\bm{F}}}. \bm{f}{\hat \varphi}$ 
 and 
 $\frac{ g_2}{2}  \left(\frac{\delta A^\dagger_{00}}{\delta {\hat \varphi}^\dagger}\right)\hat A_{00}=\frac{ g_2}{2}  (N_{00}{\hat \varphi}^\dagger {\hat \varphi} N_{00}){\hat \varphi}$, 
the  equation of motion (\ref{formal_equation_of_motion}) of the spinor BEC is
\begin{equation}
\label{equation_of_motion_spinor_BEC}
\begin{split}
i\hbar \partial_t {\hat \varphi}
=
  \left(
 -\frac{\hbar^2}{2m}\nabla^2 - \mu
 +
  g_0 {\hat \varphi}^\dagger {\hat \varphi}
 \right){\hat \varphi}
 +
 \frac{ g_1}{\hbar^2}  {\hat{\bm{F}}}. \bm{f}{\hat \varphi}
 \\
  +
\frac{ g_2}{2}  (N_{00}{\hat \varphi}^\dagger {\hat \varphi} N_{00}){\hat \varphi},
\end{split}
\end{equation}
where $\mu$ is a diagonal matrix with components $\mu_m$. 
For a point-contact interaction, the coupling strengths for a spin-1 BEC are
  $g_0  
  =
     \frac{4\pi \hbar^2}{m} \frac{ a_0 + 2 a_2  }{3}$, 
  $g_1  
  =
   \frac{4\pi \hbar^2}{m}   \frac{a_2- a_0  }{3}$ and $g_2  
  =
  0$. 
For a spin-2 BEC they are $g_0=
 \frac{4\pi \hbar^2}{m}   \frac{ 4a_2 +3 a_4  }{7}$, 
 $g_1  
  =
  \frac{4\pi \hbar^2}{m}  \frac{a_4 -a_2 }{7}$ and $g_2=
 \frac{4\pi \hbar^2}{m} \frac{7 a_0-10 a_2 + 3a_4}{7}$,
where   $a_{\cal F}$ is the $s$-wave scattering length for the binary interaction between two bosons of total spin ${\cal F}=0,2,4$ \cite{ueda_10}.

\section{Quantum lattice gas algorithm}

One may write the equation of motion  (\ref{equation_of_motion_spinor_BEC}) of a spinor BEC in   unitary evolution operator form
\begin{equation}
\label{equation_of_motion_spin_F_unitary_form}
{\hat \varphi}(\bm{r}, t+\tau) 
= 
e^{ -i \left(
-\frac{\hbar^2}{2m}\nabla^2 - \mu
 +
  g_0 {\hat \varphi}^\dagger {\hat \varphi}
+
 \frac{g_1 {\hat{\bm{F}}}\cdot \bm{f} }{\hbar^2}
  \right)\frac{\tau}{\hbar}
   }
{\hat \varphi}(\bm{r},t),
\end{equation}
where the spin-$f$ multiplet field operator is
\begin{equation}
\label{varphi_quantum_state}
{\hat \varphi} = 
\begin{pmatrix}
      {\hat \varphi}_f    &
      {\hat \varphi}_{f-1} &
      \cdots &
      {\hat \varphi}_{1-f} &
      {\hat \varphi}_{-f}  
\end{pmatrix}^\text{T}.
\end{equation}
 The method presented here is based on performing a quantum computational decomposition of the equation of motion (\ref{equation_of_motion_spin_F_unitary_form}) that employs a fermionic 2-spinor field operator ${\hat \psi}=({\hat \psi}^L, {\hat \psi}^R)^\text{T}$
\begin{subequations}
\label{psi_quantum_state}
\begin{align}
{\hat \psi}^L &= 
\begin{pmatrix}
      {\hat \psi}^L_f    &
      {\hat \psi}^L_{f-1} &
      \cdots &
      {\hat \psi}^L_{1-f} &
      {\hat \psi}^L_{-f} &
\end{pmatrix}^\text{T}
       \\
{\hat \psi}^R 
&= 
\begin{pmatrix}
      {\hat \psi}^R_f    &
      {\hat \psi}^R_{f-1} &
      \cdots &
      {\hat \psi}^R_{1-f} &
      {\hat \psi}^R_{-f} &
\end{pmatrix}^\text{T}  
\end{align}
\end{subequations}
to represent each component of the spin-$f$ multiplet (\ref{varphi_quantum_state}) as
\begin{align}
   {\hat \varphi}_m =( {\hat \psi}^L_m + {\hat \psi}^R_m )/\sqrt{2}, 
\end{align}
 for $m\in[-f, f]$.  To construct a multiple qubit and quantum gate compatible representation,   the unitary operator in (\ref{equation_of_motion_spin_F_unitary_form}) is split into kinetic energy and interaction energy parts. 
 Since the spin-$f$ field ${\hat \varphi}$ is generalized to a fermionic field operator ${\hat \psi}$,  the equation of motion that governs the ${\hat \psi}$ field operator is time-symmetrical.  The equation of motion (\ref{equation_of_motion_spin_F_unitary_form})  is modeled by an  ultraviolet unitary model 
%
\begin{align}
\nonumber
{\hat \psi}(\bm{r}, t+\tau) 
=& 
e^{ 
 -i \left(
 \frac{g_1 ({\hat{\bm{F}}}\cdot \bm{f}) \otimes \bm{1}}{\hbar^2}
  \right)
  \frac{\tau}{\hbar}
   }
e^{ 
 -i \left(
 \frac{g_2 \hat N^{00} \otimes \bm{1}}{2}
  \right)
  \frac{\tau}{\hbar}
   }
   \times
      \\
   e^{ 
 -i \bm{1}_{2f+1}\otimes\sigma_x\left(
  g_0 {\hat \varphi}^\dagger {\hat \varphi}
- \mu
  \right)
  \frac{\tau}{\hbar}
   }
   &
e^{ -i \left(
- \bm{1}_{2f+1}\otimes\sigma_x\frac{\hbar^2}{2m}\nabla^2 
\right)
  \frac{\tau}{\hbar}
   }
   {\hat \psi}(\bm{r},t),
   \label{equation_of_motion_fermionized_spin_F_unitary_form}
\end{align}
%
where $\bm{1}_{2f+1}$ is the identity matrix of size $(2f+1)\times(2f+1)$, and where
the Pauli spin matrices are
\begin{align}
\sigma_x = 
{\scriptsize
\begin{pmatrix}
  0    &   1 \\
 1     &  0
\end{pmatrix}
},
\qquad
\sigma_y = 
{\scriptsize
\begin{pmatrix}
  0    &   -i \\
 i     &  0
\end{pmatrix}
},
\qquad
\sigma_z = 
{\scriptsize
\begin{pmatrix}
  1    &   0 \\
 0     &  -1
\end{pmatrix}
}.
\end{align}
The quantum dynamics is represented in a product space of gauge groups $SO(3)\otimes SU(2)$.  The spin-$f$ representation of $SO(3)$ is $2f+1$ dimensional for the spin texture.  The spin-1/2 representation of $SU(2)\cong SO(3,1)$ is $2$ dimensional because it is a nonrelativistic representation.\footnote{A relativistic representation of SO(3,1) requires a $4$ dimensional representation, requiring $4\times 4$ Dirac matrices and a 4-spinor field.}    In this product space representation, these generators are anticommuting as they respectively reside in separate subspaces
\begin{align}
   \left[ \frac{g_1({\hat{\bm{F}}}\cdot \bm{f}) \otimes \bm{1}}{\hbar^2}
   + \frac{g_2 \hat N^{00} \otimes \bm{1}}{2}, 
    \bm{1}_{2f+1}\otimes\sigma_x\ell^2\nabla^2 \right]=0.
\end{align}
This serves as the basis for the operator splitting method used in the quantum lattice gas model. 

The fermionic field $\psi(x)$ is represented on a qubit array that encodes  spacetime as  a lattice, the fermionic field operators $\hat\psi^\dagger$ and $\hat \psi$ are represented by qubit creation and annihilation operators,  and the quantum algorithm is constructed in such a way as to reduce the particle dynamics to a sequence of unitary quantum gate operations. The desired quantum lattice gas algorithm to represent (\ref{equation_of_motion_fermionized_spin_F_unitary_form}) has second-order numerical convergence (i.e. doubling the grid resolution reduces the numerical error by one fourth for the spin-1 BEC and by over one third for the spin-2 BEC).    In short, the ansatz is to  split the quantum dynamics (\ref{equation_of_motion_spin_F_unitary_form}) into a product of unitary operators (\ref{equation_of_motion_fermionized_spin_F_unitary_form}), and then  deconstruct the unitary operators in this equation in a way that is suitable for quantum computing.  The quantum algorithm for the many-body quantum system takes the split form
\begin{equation}
\label{split_unitary_operators}
\hat \psi(\bm{r}, t+\tau)  =
 U^\text{\tiny nd}_f [\hat \psi]  U^\text{\tiny d} [\hat\psi]  U_\circ
\,
\hat\psi(\bm{r},t),
\end{equation}
 where the free particle motion  is generated by the kinetic energy  $U_\circ=e^{ -i \left(
-\sigma_x\otimes \bm{1}_{2f+1}\frac{\hbar^2}{2m}\nabla^2 
\right)
  \frac{\tau}{\hbar}
   }$, 
where the diagonal part of the nonlinear interaction is generated by the probability density $U^\text{\tiny d} [\hat\psi]
\equiv 
e^{ 
 -i \bm{1}_{2f+1}\otimes\sigma_x\left(
  g_0 {\hat \varphi}^\dagger {\hat \varphi}
- \mu
  \right)
  \frac{\tau}{\hbar}
   }
$,  where the nondiagonal part of the nonlinear interaction is generated by the spin density vector and spin-singlet density $U^\text{\tiny nd}_f [\hat\psi]\equiv e^{-i \frac{g_1 ({\hat{\bm{F}}}\cdot \bm{f}) \otimes \bm{1} }{\hbar^2}  \frac{\tau}{\hbar}} e^{ 
 -i \left(
 \frac{g_2 \hat N^{00} \otimes \bm{1}}{2}
  \right)
  \frac{\tau}{\hbar}
   }
$, and where the spin magnitude is $F = \sqrt{F_x^2 + F_y^2 + F_z^2}$.  The quantum algorithms for $U_\circ$  \cite{yepez-cpc01} and $U^\text{\tiny d} [\hat\psi]$  \cite{yepez-vahala-EPJ-09,yepez:084501,yepez:770209} are known and well tested. The quantum algorithm for $U^\text{\tiny nd}_f [\hat\psi]$ for $f=1,2$  \cite{yepez_arXiv1609.02229_cond_mat.quant_gas} is new and has performed successfully in recent numerical tests of soliton-soliton collisions and interacting non-Abelian quantum vortices.

\subsection{Free massive fermion}

The unitary quantum lattice gas algorithm for evolving a 2-spinor field
\begin{align}
\label{psi_2_spinor_field}
   {\hat \psi}(x)
=
\begin{pmatrix}
      {\hat \psi}^L(x)   \\
      {\hat \psi}^R(x)
\end{pmatrix}, 
\end{align}
where spacetime points $x=(\bm{x}, t)$ and spatial points $\bm{x}$ on a three-dimensional cubical grid \cite{yepez-cpc01}.
One can begin to construct the quantum algorithm to model a free nonrelativistic massive fermion by using a number operator 
\(
{\cal N}^\text{\tiny C} = \frac{1}{2}\left( 1-\sigma_x\right),
\)
which is idempotent $({\cal N}^\text{\tiny C})^2={\cal N}^\text{\tiny C}$.
 The unitary  operator generated by this number operator is 
\begin{equation}
\label{collision_operator}
 {\cal C}\equiv e^{i   \frac{\pi}{2} {\cal N}^\text{\tiny C}}= 1+ \left(e^{i\frac{\pi}{2}}-1\right){\cal N}^\text{\tiny C} = 
 \frac{1}{2}
 \begin{pmatrix}
  1+i    &1-i    \\
    1-i  &  1+i
\end{pmatrix},
\end{equation}
and it is applied at every point $x$ by the local map: 
${\hat \psi}'(x) = {\cal C} \,{\hat \psi}(x) \mapsto {\hat \psi}(x)$.
The displacements of the spin up ($+1$) and spin down ($-1$)  components the 2-spinor field are implemented by stream operators:
\begin{subequations}
\label{stream_operators}
\begin{align}
{\cal S}_{\Delta{\bm{x}}, 1}
& \equiv 
e^{h \Delta{\bm{x}} \cdot \bm{\nabla}}
=
 n + e^{\Delta {\bm{x}} \cdot \nabla}\,h
=
{\scriptsize
\begin{pmatrix}
  e^{\Delta{\bm{x}} \cdot \bm{\nabla}}    &    0\\
  0    &  1
\end{pmatrix}
}, 
\\
 {\cal S}_{\Delta{\bm{x}}, -1} 
 &\equiv 
e^{n \Delta{\bm{x}} \cdot \bm{\nabla}}
=
 h + e^{\Delta {\bm{x}}\cdot \nabla} \,n
=
{\scriptsize
\begin{pmatrix}
  1 &    0\\
  0    &  e^{\Delta{\bm{x}} \cdot \bm{\nabla}}   
\end{pmatrix}
},
 \end{align}
\end{subequations}
where $n=\frac{1}{2}(1-\sigma_z)$ and $h=\frac{1}{2}(1+\sigma_z)$.   These number and hole operators (used as generators in (\ref{stream_operators})) are also idempotent, i.e. $n^2=n$ and $h^2=h$.  For $\sigma=\pm 1$, the operators (\ref{stream_operators})  can be written in manifestly unitary form 
${\cal S}_{\Delta{\bm{x}}, \sigma}
\equiv 
e^{i {\cal N}^{\cal S}_\sigma  \Delta{\bm{x}} \cdot \bm{p}/\hbar}$, 
where 
${\cal N}^{\cal S}_\sigma  \equiv  \frac{1+\sigma}{2}h +\frac{1-\sigma}{2}n$, 
expressed in terms of the quantum mechanical momentum operator  $\bm{p} \equiv -i \hbar \nabla$. The unitary stream and collide operators %
 are the basic building blocks of any quantum lattice gas algorithm. 
With the appropriate boundary conditions,  the application of (\ref{stream_operators}) on  a quantum state $|\Omega\rangle$ is guaranteed to conserve the total number density 
$\int d^3x\, \langle \Omega|{\hat \psi}^\dagger({\bm{x}}){\hat \psi}({\bm{x}}) |\Omega\rangle$. 
Consider the product operator 
\begin{subequations}
\label{interleaved_operator}
\begin{align}
I_{\bm{x}\sigma} 
&= 
 {\cal S}_{-\Delta {\bm{x}},\sigma}  {\cal C}^\dagger  {\cal S}_{\Delta {\bm{x}},\sigma}  {\cal C}
 \\
&\stackrel{(\ref{collision_operator})}{\stackrel{(\ref{stream_operators})}{=}}
 \frac{1}{2}
 \begin{pmatrix}
1+ e^{-\sigma\Delta\bm{x}\cdot \nabla}      &   -i+ i\,e^{-\sigma\Delta\bm{x}\cdot \nabla}  \\
 - i+ i\, e^{\sigma\Delta\bm{x}\cdot \nabla}     &  1+ e^{\sigma\Delta\bm{x}\cdot \nabla}
\end{pmatrix}.
\end{align}
\end{subequations}
The ordering of operators in (\ref{interleaved_operator}) is not unique since $[{\cal S}_{-\Delta {\bm{x}},\sigma}  {\cal C}^\dagger , {\cal S}_{\Delta {\bm{x}},\sigma}  {\cal C}]=0$. One  defines a symmetrized  operator 
$U_{\bm{x}\sigma}
\equiv 
I_{\bm{x}\sigma}I_{\bm{x}\bar\sigma}$, 
where  the identity $\cosh z - \frac{1}{2}\sinh z = 1 - 2 \sinh^4(z/2)$.  A suitable evolution operator to model a  quantum particle's motion in  three spatial dimensions may be constructed by  a product of such
fully symmetrized operators, one  for each orthogonal Cartesian direction
 \begin{equation}
\label{basic_quantum_algorithm_diagonal}
U_\circ 
\equiv  U_{\bm{z}} U_{\bm{y}} U_{\bm{x}} .
\end{equation}

Only the spinor field $\psi({\bm{x}},t)$ is stored on a computer simulation at one time--the method is a time-explicit method.  Yet, from a theoretical point of view, the  many-body state operator of the total entire system is actually a tensor product of all the 2-spinor field operators over all  the grid points
\begin{align}
   \hat \Psi(t) \equiv \bigotimes_{\bm{x}\in \text{grid}}\hat\psi({\bm{x}},t).
\end{align}
Likewise, the system's evolution operator is a tensor product over all the local unitary operators
\begin{align}
\label{basic_quantum_algorithm_diagonal_multiplet}
   U_\circ^\text{gas} &=
  \bigotimes_{\bm{x}\in \text{grid}} U_\circ ,
   \end{align}
which is a matrix of size $2^{2V}\times 2^{2V}$, where volume of the grid is $V=L^3$ for grid size $L$ with two qubits per point encoding $\psi$ on the qubit array.  
The  system evolution equation for a quantum gas is
\begin{align}
\label{Quantum_gas_system_evolution}
   \hat\Psi(t+\tau)\stackrel{(\ref{basic_quantum_algorithm_diagonal_multiplet})}{=}   U_\circ^\text{gas}  \hat\Psi(t).    
\end{align}
$U_\circ^\text{gas} $ encodes the basic  algorithm to model a quantum gas of particles confined to a lattice \cite{yepez-cpc01}.  

The quantum system governed by (\ref{basic_quantum_algorithm_diagonal_multiplet}) is  a quantum lattice gas, which in this case is a quantum gas of particles confined to a spacetime lattice.  The qubit array is arranged in a cubical grid.  The system evolution equation (\ref{Quantum_gas_system_evolution}) can be expressed in differential point form as the equation of motion is
\begin{align}
\label{Yepez_equation}
   i\hbar \partial_t \hat\psi(x) = 
-
\sigma_x
\frac{\hbar^2}{2m} \nabla^2  
\hat\psi(x), 
 \end{align}
which is the nonrelativistic limit of the Weyl-Dirac equation for a free fermion in the chiral representation.  That is, (\ref{Quantum_gas_system_evolution}) is used to numerically represent (\ref{Yepez_equation}) on a qubit array.   Defining the massive bosonic (pairing) field as 
\begin{align}
   {\hat \varphi}(x) =\left. \frac{1}{\sqrt{2}}\middle( \hat \psi^L(x) + \hat \psi^R(x) \right),
\end{align}
the equation of motion for the $\hat \varphi$ operator is
\begin{equation}
\label{free_boson_equation}
 i\hbar \partial_t {\hat \varphi}(x) = 
-
\frac{\hbar^2}{2m} \nabla^2  
{\hat \varphi}(x).
\end{equation}
Let  $|\Omega^{(1)}\rangle$ denote a single particle quantum state.  
 In the 1-body sector of the Hilbert space, the expectation value of (\ref{free_boson_equation}) with respect to $|\Omega^{(1)}\rangle$ becomes the well known Schroedinger wave equation for a free quantum particle with wave function $\varphi(x) = \langle\Omega^{(1)}|{\hat \varphi}(x)|\Omega^{(1)}\rangle$ 
\begin{equation}
 i\hbar \partial_t  \varphi(x) = 
-
\frac{\hbar^2}{2m} \nabla^2  
 \varphi(x).
\end{equation}

Understanding the behavior of quantum lattice gases on supercomputers allows us to better understand ultracold quantum gases, and this work bridges the gap between  quantum computing and analog quantum simulation.  The connection between quantum computing and ultracold quantum gases  becomes more apparent in the following representations of BECs.  Yet, to model a BEC the $\hat \varphi^4$ nonlinear terms in (\ref{diagonal_energy_functional}) and (\ref{nondiagonal_energy_functional}) must be represented in terms of separate unitary operators that multiply the righthand side of (\ref{Quantum_gas_system_evolution}).

\subsection{Scalar BEC model}

The number density operator for the bosonic field in the quantum lattice gas  is
\begin{align}
   \hat \rho(x) 
   &= \hat\varphi^\dagger(x)\hat \varphi(x) 
   \\
   &=
   \left.
   \frac{1}{2}
   \middle(
   \hat\psi^{\text{\tiny L}\dagger}(x) + \hat\psi^{\text{\tiny R}\dagger}(x)
\middle)
\middle(
   \hat\psi^{\text{\tiny L}}(x) + \hat\psi^{\text{\tiny R}}(x)
   \right)
   .
   \end{align}
To model a scalar BEC, one can add a nonlinear interaction potential ${\cal U}(
\hat\rho)$ by representing it as the time-component of the 4-vector potential in the Abelian U(1) gauge group.  ${\cal U}(\hat\rho)=g_0\hat\rho/2$ in the Hartree-Fock approximation,  where the coupling parameter is   $g_0=\hbar^2 4\pi a/m + \cdots$ for scattering length $a$.   The unitary evolution operator for the scalar BEC is 
 \begin{equation}
\label{basic_quantum_algorithm_diagonal}
U^\text{\tiny BEC}[\hat\psi] \equiv  e^{-i  \frac{{\cal U}(\hat\rho)-\mu}{m_\circ \ell^2/\tau^2} \sigma_x}   U_\circ^\text{gas} ,
\end{equation}
 where the particle mass is $m = m_\circ/2$, and the chemical potential is $\mu$.    
$U^\text{\tiny BEC}[\hat\psi]$ represents the basic quantum lattice gas algorithm  \cite{yepez-vahala-EPJ-09,yepez:084501,yepez:770209}  used to numerically model a scalar BEC.  The equation of motion for  $\hat\varphi$ is
\begin{equation}
\label{Gross_Pitaevskii_equation_operator_form}
i \hbar\partial_t \hat\varphi(x) = - \frac{\hbar^2}{2m}\nabla^2\hat \varphi(x) + (g  \hat\varphi^\dagger(x) \hat \varphi(x)  -\mu)\,\hat\varphi(x).
\end{equation} 
The expectation value of (\ref{Gross_Pitaevskii_equation_operator_form}), in the mean-field limit, becomes the well known Gross-Pitaevskii (GP) equation \cite{JMathPhys.1963.4.195,JETP.1961.2.451} for a spin-0 BEC superfluid
\begin{equation}
\label{Gross_Pitaevskii_equation_again}
i \hbar\partial_t \varphi(x) = - \frac{\hbar^2}{2m}\nabla^2 \varphi(x) + (g  \varphi^\ast(x)  \varphi(x)  -\mu)\,\varphi(x).
\end{equation} 
%

\subsection{Spinor BEC models}

To model a spinor BEC, the quantum state (\ref{psi_2_spinor_field}) is given a hyperfine spin-indexed 4-spinor field
\begin{align}
\label{psi_2_spinor_field_for_hyperfine_level_m}
   \hat{\psi}_m(x)
=
\begin{pmatrix}
      \hat{\psi}^L_m(x)   \\
      \hat{\psi}^R_m(x)
\end{pmatrix},
\end{align}
where the hyperfine level is $m\in[-f, f]$ for a spin-$f$ BEC.  
In this way, the spin-1/2 field $\hat{\psi}$ is promoted to multiplet fermionic field (\ref{psi_quantum_state}).  That is, the fermionic field in the Zeeman hyperfine manifold can be expressed as a direct sum
\begin{align}
\label{psi_direct_sum_in_Zeeman_manifold}
   \hat{\psi} &
=
\bigoplus  _{m=-f}^f \hat{\psi}_m
=
\begin{pmatrix}
      \hat{\psi}^L_f    \\
      \hat{\psi}^R_f \\
      \vdots\\
      \hat{\psi}^L_{-f}    \\
      \hat{\psi}^R_{-f} \\        
\end{pmatrix},
\end{align}
which is a multiplet probability amplitude field with $2(2f+1)$ components. 
 For a spin-$f$ multiplet field, the diagonal part of the evolution operator for the spinor BEC can also be expressed as a direct sum 
\begin{align}
\label{basic_quantum_algorithm_diagonal_multiplet}
   U^\text{\tiny d} [\hat{\psi}] &=
   \bigoplus_{m=-f}^f  U^\text{\tiny BEC}[\hat{\psi}_m]  ,
   \end{align}
which is a matrix of size $2(2f+1)\times 2(2f+1)$.  
In the low-energy and low-momentum limits, the resulting nonrelativistic equation of motion for each Zeeman level is 
\begin{equation}
\label{equation_of_motion_spinor_BEC_diagonal_part}
i\hbar \partial_t \hat{\psi}_m
=
 \sigma_x \left(
 -\frac{\hbar^2}{2m}\nabla^2 - \mu
 +
  g_0 \hat{\varphi}^\dagger \hat{\varphi}
 \right)\hat{\psi}_m,
\end{equation}
 where $\hat{\psi}_m = \begin{pmatrix}
     \hat{\psi}^L_m    & \hat{\psi}^R_m       
\end{pmatrix}^\text{T}$ for $m\in[-f, f]$.   Equivalently, this can be written for the full fermionic field (\ref{psi_direct_sum_in_Zeeman_manifold}) as a tensor product over the levels in the Zeeman manifold  
\begin{equation}
\label{equation_of_motion_spinor_BEC_diagonal_part_for_psi}
i\hbar \partial_t \hat{\psi}
=
\bm{1}_{2f+1}\otimes \sigma_x \left(
 -\frac{\hbar^2}{2m}\nabla^2 - \mu
 +
  g_0 \varphi^\dagger \varphi
 \right)\hat{\psi}.
\end{equation}
Finally, to model the spinor GP equations (\ref{equation_of_motion_spinor_BEC}) for a non-Abelian superfluid, nonlinear unitary interactions are appropriately added to the product (\ref{basic_quantum_algorithm_diagonal_multiplet}) to model, for example, either an $f=1$ or $f=2$ spinor BEC. The quantum algorithmic procedures for doing this for the spin-1 and spin-2 cases are described next.

\subsubsection{$f=1$ spinor BEC model}

A spin $f=1$ matrix representation of the SU(2) Lie algebra $[f_i, f_j]= i \hbar \epsilon_{ijk} f_k$ is
\begin{equation}
f_x =
\frac{\hbar}{\sqrt{2}}
{\scriptsize
\begin{pmatrix}
  0    & 1 & 0    \\
  1    & 0 & 1 \\
  0  & 1 & 0 
\end{pmatrix}
},\;
f_y =
\frac{\hbar}{\sqrt{2}}
{\scriptsize
\begin{pmatrix}
  0    & -i & 0    \\
  i    & 0 & -i\\
  0  & i & 0 
\end{pmatrix}
},\;
f_z =
\hbar
{\scriptsize
\begin{pmatrix}
  1    & 0 & 0    \\
  0    & 0 & 0 \\
  0  & 0 & -1 
\end{pmatrix}
}.
\end{equation}
Since the cube of the generator is proportional to the generator itself
$( {\hat{\bm{F}}}\cdot \bm{f} )^3 = (F_x^2+ F_y^2 + F_z^2) {\hat{\bm{F}}}\cdot \bm{f} = F^2 {\hat{\bm{F}}}\cdot \bm{f}$, 
the nondiagonal interaction  is generated by a  tri-idempotent number operator ($\bm{N}^2\neq \bm{N}$ and $\bm{N}^3= \bm{N}$)
\begin{subequations}
\begin{eqnarray}
\bm{N} 
&\equiv &
\frac{{\hat{\bm{F}}}\cdot \bm{f}}{\hbar}
=
\frac{1}{F}
{\scriptsize
\begin{pmatrix}
    F_z  &  \frac{F_x - i F_y}{\sqrt{2}} & 0    \\
     \frac{F_x + i F_y}{\sqrt{2}}   &  0 &  \frac{F_x - i F_y}{\sqrt{2}} \\
     0 &  \frac{F_x + i F_y}{\sqrt{2}} & - F_z
\end{pmatrix}
}
\\
\bm{N}^2
&=&
\frac{1}{F^2}
{\scriptsize
\begin{pmatrix}
    \frac{F_x^2+F_y^2}{2} + F_z^2  &  \frac{(F_x - i F_y)F_z}{\sqrt{2}} & \frac{(F_x-iF_y)^2}{2}    \\
    \frac{(F_x+iF_y)^2}{2} &  F_x^2 + F_y^2 &  - \frac{(F_x - i F_y)F_z}{\sqrt{2}}  \\
      \frac{(F_x+iF_y)^2}{2}   &   - \frac{(F_x + i F_y)F_z}{\sqrt{2}}  &     \frac{F_x^2+F_y^2}{2} + F_z^2 
\end{pmatrix}
}.
\qquad
\;
\end{eqnarray}
\end{subequations}
For a spin-1 BEC,  $U^\text{\tiny nd}_{f=1} [\hat{\varphi}]$ on the righthand side of (\ref{split_unitary_operators})  can  be analyticaly expanded to all orders
\begin{subequations}
\label{nondiagonal_equation_of_motion_spin_1_unitary_form_analytical_expansion}
\begin{eqnarray}
\nonumber
U^\text{\tiny nd}_{f=1}  [\hat{\varphi}]
\!\!\!
& =&
\!\!\!
{\scriptsize 
 \left.
  1 + \middle( \cos\left(\frac{g_1 F\tau}{\hbar^2}\right) -1 \middle)
\hat{\bm{N}}^2
 - i \sin\left(\frac{g_1 F\tau}{\hbar^2}\right) 
\hat{\bm{N}}
\right.
}
\\
\\
\nonumber
=1& -&
\frac{g_1^2 \tau^2}{2\hbar^4}
\hat{\bm{N}}^2
\text{sinc}^2
\frac{g_1 F\tau}{2\hbar^2}
 - i 
 \frac{g_1 \tau}{\hbar^2}
 \hat{\bm{N}}
\text{sinc}
\frac{g_1 F\tau}{2\hbar^2}
.
\\
\label{nondiagonal_equation_of_motion_spin_1_unitary_form_analytical_expansion_b}
\end{eqnarray}
\end{subequations}
Therefore, the equation of motion for the spin-1 BEC is modeled by the quantum algorithm (expressed in split unitary operator form) 
\begin{equation}
\label{spin_1_BEC_quantum_algorithm}
\hat{\psi}(\bm{x}, t+\tau) 
\stackrel{(\ref{split_unitary_operators})}{=}
U^\text{\tiny nd}_{f=1}  [\hat{\psi}] 
U^\text{\tiny d} [\hat{\psi}] 
\,
\hat{\psi}(\bm{x},t),
\end{equation}
where 
\begin{align}
   U^\text{\tiny nd}_{f=1}  [\hat{\psi}] =U^\text{\tiny nd}_{f=1}  [\hat{\varphi}] \otimes \bm{1}. 
\end{align}
%

\subsubsection{$f=2$ spinor BEC model}

A spin $f=2$ matrix representation of the SU(2) Lie algebra $[f_i, f_j]= i \hbar \epsilon_{ijk} f_k$ is
\begin{widetext}
\begin{equation}
\label{standard_spin_2_representation}
f_x 
=
\hbar
{\scriptsize
\begin{pmatrix}
  0    & 1 & 0   & 0 & 0  \\
  1    & 0 & \sqrt{\frac{3}{2}} & 0 & 0 \\
  0  & \sqrt{\frac{3}{2}}  & 0 & \sqrt{\frac{3}{2}}  & 0\\
  0 & 0 & \sqrt{\frac{3}{2}} & 0 & 1\\
  0 & 0 & 0 & 1 & 0 \\ 
\end{pmatrix}
}
\qquad
f_y 
=
\hbar
{\scriptsize
\begin{pmatrix}
  0    & -i & 0   & 0 & 0  \\
  i    & 0 & -i\sqrt{\frac{3}{2}} & 0 & 0 \\
  0  & i \sqrt{\frac{3}{2}}  & 0 & -i\sqrt{\frac{3}{2}}  & 0\\
  0 & 0 & i\sqrt{\frac{3}{2}} & 0 & -i\\
  0 & 0 & 0 & i & 0 \\ 
\end{pmatrix}
}
\qquad
f_z 
=
\hbar
{\scriptsize
\begin{pmatrix}
  2    & 0 & 0  & 0 & 0  \\
  0    & 1 & 0 & 0 & 0 \\
  0    & 0 & 0 & 0 & 0 \\
  0  & 0 & 0 & -1 & 0 \\
  0 & 0 & 0 & 0 & -2\\
\end{pmatrix}
}.
\end{equation}
The number operator generating the evolution is
\begin{equation}
\label{spin_F_2_number_operator}
\bm{N} \equiv \frac{{\hat{\bm{F}}}\cdot \bm{f}}{\hbar}
=
\frac{1}{F}
{\scriptsize
\begin{pmatrix}
  2 F_z    & F_x -i F_y& 0   & 0 & 0  \\
  F_x + i F_y   & F_z  & \sqrt{\frac{3}{2}} (F_x -i F_y)& 0 & 0 \\
  0  & \sqrt{\frac{3}{2}} ( F_x + i  F_y) & 0 & \sqrt{\frac{3}{2}} (F_x-i  F_y)  & 0\\
  0 & 0 & \sqrt{\frac{3}{2}} (F_x+i  F_y) & -F_z  & F_x-i  F_y \\
  0 & 0 & 0 & F_x +i  F_y & -2 F_z  \\ 
\end{pmatrix}
}
.
\qquad
\end{equation}
Since the number operator (\ref{spin_F_2_number_operator}) is neither idempotent nor tri-idempotent, there is no closed-form expansion of  $U^\text{\tiny nd}_{f=2}  [\hat{\varphi}]$
to all orders using a generalized Euler identity. 
Nevertheless,  the spin-density part of the nondiagonal evolution operator $U^\text{\tiny nd}_{f=2}  [{\hat{\bm{F}}},\hat{\varphi}]=e^{-i \left(
\frac{g_1 F \tau}{\hbar} \right)
\left(
\frac{\hat{{\hat{\bm{F}}}}\cdot \bm{f}}{\hbar}
\right)
/\hbar
}
$ expanded to lowest-order
\begin{equation}
\label{nondiagonal_equation_of_motion_spin_2_unitary_matrix_form_analytical_expansion}
U^\text{\tiny nd}_{f=2}  [{\hat{\bm{F}}},\hat{\varphi}]
  =
\left.
{\scriptsize
\begin{pmatrix}
   1   & 0 & 0    & 0 & 0\\
     0 &  1 & 0 & 0 &0 \\
     0 & 0 & 1 & 0 & 0 \\
   0   & 0 & 0    & 1 & 0\\
   0   & 0 & 0    & 0 & 1\\
\end{pmatrix}
}
 - i 
\frac{g_1 \tau}{\hbar^2}
{\scriptsize
\begin{pmatrix}
  2 F_z    & F_x -i F_y& 0   & 0 & 0  \\
  F_x + i F_y   & F_z  & \sqrt{\frac{3}{2}} (F_x -i F_y)& 0 & 0 \\
  0  & \sqrt{\frac{3}{2}} ( F_x + i  F_y) & 0 & \sqrt{\frac{3}{2}} (F_x-i  F_y)  & 0\\
  0 & 0 & \sqrt{\frac{3}{2}} (F_x+i  F_y) & -F_z  & F_x-i  F_y \\
  0 & 0 & 0 & F_x +i  F_y & -2 F_z  \\ 
\end{pmatrix}
}
\right.
+\cdots
\end{equation}
can be used for analytical matching purposes. 
That is, although there are no known  spin-2 representations of SU(2) that can be exactly summed to infinite order,  
 it is possible
  to perform an infinite-order expansion that is an analytical continuation of  $U^\text{\tiny nd}_{f=2}  [{\hat{\bm{F}}},\hat{\varphi}]$ that matches (\ref{nondiagonal_equation_of_motion_spin_2_unitary_matrix_form_analytical_expansion}) at lowest order. 
    Using such an analytical continuation, one is free to choose the value of $g_1$ to be order unity for modeling  nonperturbative quantum matter.  Consider the set of number operators, rendered in the   spin-1/2 subspaces $m=2,1$ and $m=-1,-2$ 
\begin{equation}
\label{standard_spin_2_representation_parallel}
f_x^{({1}/{2})} 
=
\frac{\hbar}{2}
{\scriptsize
\begin{pmatrix}
  0    & 1 & 0   & 0 & 0  \\
  1    & 0 & 0 & 0 & 0 \\
  0  & 0  & 0 & 0  & 0\\
  0 & 0 & 0 & 0 & 1\\
  0 & 0 & 0 & 1 & 0 \\ 
\end{pmatrix}
}
\qquad
f_y^{({1}/{2})} 
=
\frac{\hbar}{2}
{\scriptsize
\begin{pmatrix}
  0    & -i & 0   & 0 & 0  \\
  i    & 0 & 0 & 0 & 0 \\
  0  & 0  & 0 & 0  & 0\\
  0 & 0 & 0 & 0 & -i\\
  0 & 0 & 0 & i & 0 \\ 
\end{pmatrix}
}
\qquad
f_z^{({1}/{2})} 
=
\frac{\hbar}{2}
{\scriptsize
\begin{pmatrix}
  1    & 0 & 0  & 0 & 0  \\
  0    & -1& 0 & 0 & 0 \\
  0    & 0 & 0 & 0 & 0 \\
  0  & 0 & 0 & 1 & 0 \\
  0 & 0 & 0 & 0 & -1\\
\end{pmatrix}
}
\end{equation}
and in the  spin-1 subspace $m=1,0,-1$
\begin{equation}
\label{standard_spin_2_representation_perpendicular}
f_x^{(1)} 
=
\frac{\hbar}{\sqrt{2}}
{\scriptsize
\begin{pmatrix}
  0    & 0 & 0   & 0 & 0  \\
  0    & 0 & 1 & 0 & 0 \\
  0  & 1  & 0 & 1  & 0\\
  0 & 0 & 1 & 0 & 0\\
  0 & 0 & 0 & 0 & 0 \\ 
\end{pmatrix}
}
\qquad
f_y^{(1)} 
=
\frac{\hbar}{\sqrt{2}}
{\scriptsize
\begin{pmatrix}
  0    & 0 & 0   & 0 & 0  \\
  0    & 0 & -i & 0 & 0 \\
  0  & i  & 0 & -i  & 0\\
  0 & 0 & i & 0 & 0\\
  0 & 0 & 0 & 0 & 0 \\ 
\end{pmatrix}
}
\qquad
f_z^{(1)} 
=
\hbar
{\scriptsize
\begin{pmatrix}
  0    & 0 & 0  & 0 & 0  \\
  0    & 1 & 0 & 0 & 0 \\
  0    & 0 & 0 & 0 & 0 \\
  0  & 0 & 0 & -1 & 0 \\
  0 & 0 & 0 & 0 & 0\\
\end{pmatrix}
}.
\end{equation}
Spin-2 spin-density interaction dynamics may be generated by a spin-1/2 nonlinear tri-idempotent number operator ($\bm{N}^2_{({1}/{2})}\neq \bm{N}_{({1}/{2})}$ and $\bm{N}_{({1}/{2})}^3= \bm{N}_{({1}/{2})}$)
\begin{equation}
\bm{N}_{({1}/{2})}[{\hat{\bm{F}}}_{({1}/{2})}]
\equiv 
\frac{2{{\hat{\bm{F}}}_{({1}/{2})}}\cdot \bm{f}_{({1}/{2})}}{\hbar F_{({1}/{2})}}
=
\frac{1}{F_{({1}/{2})}}
{\scriptsize
\begin{pmatrix}
    F_z^{({1}/{2})}  &  F_x^{({1}/{2})} - i F_y^{({1}/{2})} & 0 & 0 & 0     \\
       F_x^{({1}/{2})} + i F_y^{({1}/{2})}    & -F_z^{({1}/{2})} & 0 & 0 & 0 \\
    0  &  0 &  0 & 0 &0 \\
       0    & 0 & 0 & F_z^{({1}/{2})} & F_x^{({1}/{2})}- i F_y^{({1}/{2})} \\
     0 & 0 &  0& F_x^{({1}/{2})} + i F_y^{({1}/{2})}  & - F_z^{({1}/{2})}
\end{pmatrix}
},
\end{equation}
and a spin-1 nonlinear tri-idempotent number operator ($\bm{N}^2_{(1)}\neq \bm{N}_{(1)}$ and $\bm{N}_{(1)}^3= \bm{N}_{(1)}$)
\begin{equation}
\bm{N}_{(1)} [{\hat{\bm{F}}}_{(1)}]
\equiv 
\frac{{{\hat{\bm{F}}}_{(1)}}\cdot \bm{f}_{(1)}}{\hbar F_{(1)}}
=
\frac{1}{F_{(1)}}
{\scriptsize
\begin{pmatrix}
  0 & 0 & 0 & 0 & 0 \\ 
  0 &   F_z^{(1)}  &  \frac{F_x^{(1)} - i F_y^{(1)}}{\sqrt{2}} & 0   & 0  \\
   0 &   \frac{F_x^{(1)} + i F_y^{(1)}}{\sqrt{2}}   &  0 &  \frac{F_x^{(1)} - i F_y^{(1)}}{\sqrt{2}} & 0 \\
   0&   0 &  \frac{F_x^{(1)} + i F_y^{(1)}}{\sqrt{2}} & - F_z^{(1)} & 0 \\
       0 & 0 & 0 & 0 & 0 \\ 
\end{pmatrix}
}.
\end{equation}
One can employ a  unitary decomposition  of the spin-2 spin-density dynamics as a product of spin-1/2 and spin-1 dynamics. i.e.
$e^{-i \left(
\frac{g_1 F \tau}{\hbar} \right)
\left(
\frac{\hat{{\hat{\bm{F}}}}\cdot \bm{f}}{\hbar}
\right)
/\hbar
}
\approx
e^{-i \left(
\frac{g_1 F_{({1}/{2})} \tau}{\hbar} \right)
\bm{N}_{({1}/{2})}[{\hat{\bm{F}}}_{({1}/{2})}]
/\hbar
}
e^{-i \left(
\frac{g_1 F_{(1)} \tau}{\hbar} \right)
\bm{N}_{(1)}[{\hat{\bm{F}}}_{(1)}]
/\hbar
}$. 
So to match (\ref{nondiagonal_equation_of_motion_spin_2_unitary_matrix_form_analytical_expansion}), an expansion of $U^\text{\tiny nd}_{f=2}  [{\hat{\bm{F}}},\hat{\varphi}]$ about $g_1$ may be written as
\begin{subequations}
\label{nondiagonal_equation_of_motion_spin_2_unitary_matrix_form_analytical_continuation}
\begin{eqnarray}
U^\text{\tiny nd}_{f=2}  [{\hat{\bm{F}}},\hat{\varphi}]
&\approx&
e^{-i \left(
\frac{g_1 F_{({1}/{2})} \tau}{\hbar^2} \right)
\bm{N}_{({1}/{2})}[{\hat{\bm{F}}}_{({1}/{2})}]
}
e^{-i \left(
\frac{g_1 F_{(1)} \tau}{\hbar^2} \right)
\bm{N}_{(1)}[{\hat{\bm{F}}}_{(1)}]
}
\\
 & =&
\left.
{\scriptsize
\begin{pmatrix}
   1   & 0 & 0    & 0 & 0\\
     0 &  1 & 0 & 0 &0 \\
     0 & 0 & 1 & 0 & 0 \\
   0   & 0 & 0    & 1 & 0\\
   0   & 0 & 0    & 0 & 1\\
\end{pmatrix}
}
 - i 
\frac{g_1 \tau}{\hbar^2}
{\scriptsize
\begin{pmatrix}
  2 F_z    & F_x -i F_y& 0   & 0 & 0  \\
  F_x + i F_y   & F_z  & \sqrt{\frac{3}{2}} (F_x -i F_y)& 0 & 0 \\
  0  & \sqrt{\frac{3}{2}} ( F_x + i  F_y) & 0 & \sqrt{\frac{3}{2}} (F_x-i  F_y)  & 0\\
  0 & 0 & \sqrt{\frac{3}{2}} (F_x+i  F_y) & -F_z  & F_x-i  F_y \\
  0 & 0 & 0 & F_x +i  F_y & -2 F_z  \\ 
\end{pmatrix}
}
\right.
+\cdots
,
\end{eqnarray}
\end{subequations}
where continuation in the small $g_1$ regime is ensured by  choosing  
${\hat{\bm{F}}}_{({1}/{2})}  = (F_x,  F_y , 2 F_z)$ and ${\hat{\bm{F}}}_{(1)}  =   (\sqrt{3} F_x, \sqrt{3} F_y , 3 F_z)$. 
 This nonperturbative analytical continuation becomes exactly computable by employing  generalized Euler identities
\begin{subequations}
\label{closed_form_F_1_F_2_Euler_identities}
\begin{eqnarray}
U^\text{\tiny nd}_{f=1/2}  [{\hat{\bm{F}}}_{({1}/{2})},\hat{\varphi}]
=
e^{-i \left(
\frac{g_1 F_{({1}/{2})} \tau}{\hbar^2} \right)
\bm{N}_{({1}/{2})}
}
&=&
1
-
\frac{g_1^2 \tau^2}{2\hbar^4}
\hat{\bm{N}}_{({1}/{2})}^2
\text{sinc}^2
\frac{g_1 F_{({1}/{2})}\tau}{2\hbar^2}
 - i 
 \frac{g_1 \tau}{\hbar^2}
 \hat{\bm{N}}_{({1}/{2})}
\text{sinc}
\frac{g_1 F_{({1}/{2})}\tau}{2\hbar^2}
\\
U^\text{\tiny nd}_{f=1}  [{\hat{\bm{F}}}_{(1)} ,\hat{\varphi}]
=
e^{-i \left(
\frac{g_1 F_{(1)} \tau}{\hbar^2} \right)
\bm{N}_{(1)}
}
&=&
1
-
\frac{g_1^2 \tau^2}{2\hbar^4}
\hat{\bm{N}}_{(1)}^2
\text{sinc}^2
\frac{g_1 F_{(1)}\tau}{2\hbar^2}
 - i 
 \frac{g_1 \tau}{\hbar^2}
 \hat{\bm{N}}_{(1)}
\text{sinc}
\frac{g_1 F_{(1)}\tau}{2\hbar^2}
.
\end{eqnarray}
\end{subequations}
So  the closed-form infinite expansions (\ref{closed_form_F_1_F_2_Euler_identities}) can be used to handle the  $g_1\sim 1$ regime while retaining  unitarity of the matrix representation of 
\begin{align}
\label{nondiagonal_spin_density_evolution_operator}
   U^\text{\tiny nd}_{f=2}  [{\hat{\bm{F}}},\hat{\varphi}]  \approx U^\text{\tiny nd}_{f=1/2}  [{\hat{\bm{F}}}_{({1}/{2})},\hat{\varphi}] U^\text{\tiny nd}_{f=1}  [{\hat{\bm{F}}}_{(1)} ,\hat{\varphi}].  
   \end{align}
For the $f=2$ case, let us define the matrix $N^{00}$ 
\begin{align}
\label{N_00_matrix}
   N^{00} 
   &=
   \frac{1}{\sqrt{5}}
   {\scriptsize
   \begin{pmatrix}
   0   & 0 & 0 & 0 & 1    \\
   0   & 0 & 0 & -1 & 0    \\
   0   & 0 & 1 & 0 & 0    \\
   0   & -1 & 0 & 0 & 0    \\
   1   & 0 & 0 & 0 & 0   \\
\end{pmatrix}
}
     ,
\end{align}
which is an involution operator.  In turn, we may calculate the value of $\hat A_{00}(\bm{x},\bm{x}')$ as
\begin{subequations}
\label{A_00_form_1_mf}
\begin{align}
\label{A_00_form_1_a}
    \hat A_{00}(\bm{x},\bm{x}') 
   &=  \hat{\varphi}(\bm{x}) N^{00}  \hat{\varphi}(\bm{x}')
   \\
\label{A_00_form_1_mf_b}
&=
\left.
\frac{1}{\sqrt{5}}
\middle(
  \hat{\varphi}_{2}(\bm{x})\hat{\varphi}_{-2}(\bm{x}')
 +
   \hat{\varphi}_{-2}(\bm{x})\hat{\varphi}_{2}(\bm{x}')
-
  \hat{\varphi}_{1}(\bm{x}) \hat{\varphi}_{-1}(\bm{x}')
-
  \hat{\varphi}_{-1}(\bm{x}) \hat{\varphi}_{1}(\bm{x}')
+
 \hat{\varphi}_0(\bm{x})\hat{\varphi}_0(\bm{x}')
\right)
.
\end{align}
\end{subequations}
If we define the 2-point spin-singlet density operator   $\hat N^{00}(\bm{x},\bm{x}')$ implicitly as
\begin{align}
\frac{\delta( \hat A_{00}^\dagger(\bm{x},\bm{x}')   \hat A_{00}(\bm{x},\bm{x}')  )  }{\delta  \hat\varphi^\dagger(\bm{x}')}
&
\equiv
\hat N^{00}(\bm{x},\bm{x}') 
\,
 \hat{\varphi}(\bm{x})
,
\end{align}
then  this number operator may be formally written  as
\begin{equation}
\label{N00_operator_formal_identity_mf}
\hat N^{00}(\bm{x},\bm{x}') 
=
\frac{\delta  \hat A_{00}^\dagger(\bm{x},\bm{x}') }{\delta  \hat\varphi^\dagger(\bm{x}') }
\cdot
\frac{  \hat A_{00}(\bm{x},\bm{x}')}{ \hat{\varphi}(\bm{x})},
\end{equation}
since $\delta  \hat A_{00}/\delta  \hat{\varphi}^\dagger\stackrel{(\ref{A_00_form_1_a})}{=}0$. 
Furthermore, since ${\delta  \hat A_{00}^\dagger }/{\delta  \hat{\varphi}^\dagger}= N^{00}  \hat{\varphi}^\dagger$, it is possible to evaluate the formal expression of $\hat N^{00}$ by converting it to matrix form 
\begin{subequations}
\begin{align}
\label{N_00_operator_matrix_form_mf}
 \hat N^{00}(\bm{x},\bm{x}')
 &=
 (N^{00}  \hat{\varphi}^\dagger(\bm{x}))
 \cdot
 ( \hat{\varphi}(\bm{x}') N^{00})
 \\
    &=
    \frac{1}{\sqrt{5}}
\begin{pmatrix}
      \hat{\varphi}_{-2}^\dagger(\bm{x})
          \\
      -\hat{\varphi}_{-1}^\dagger(\bm{x})
\\
      \hat{\varphi}_{0}^\dagger(\bm{x})
\\
      -\hat{\varphi}_{1}^\dagger(\bm{x})
\\
      \hat{\varphi}_{2}^\dagger(\bm{x})
\end{pmatrix}
\cdot
    \frac{1}{\sqrt{5}}
\begin{pmatrix}
      \hat{\varphi}_{-2}(\bm{x}')
          &
      -\hat{\varphi}_{-1}(\bm{x}')
&
      \hat{\varphi}_{0}(\bm{x}')
&
      -\hat{\varphi}_{1}(\bm{x}')
&
      \hat{\varphi}_{2}(\bm{x}')
\end{pmatrix}
       \\
\label{N_00_operator_matrix_form_c}
    &=
    \frac{1}{5}
    \begin{pmatrix}
          \hat\varphi_{-2}^\dagger(\bm{x})    \hat\varphi_{-2}(\bm{x}') 
          &
          -\hat\varphi_{-2}^\dagger(\bm{x})    \hat\varphi_{-1}(\bm{x}')           &
          &
          \hat\varphi_{-2}^\dagger(\bm{x})    \hat\varphi_{0}(\bm{x}') 
          &
         - \hat\varphi_{-2}^\dagger(\bm{x})    \hat\varphi_{1}(\bm{x}') 
          &
          \hat\varphi_{-2}^\dagger(\bm{x})    \hat\varphi_{2}(\bm{x}') 
\\
       - \hat\varphi_{-1}^\dagger(\bm{x})    \hat\varphi_{-2}(\bm{x}') 
          &
          \hat\varphi_{-1}^\dagger(\bm{x})    \hat\varphi_{-1}(\bm{x}')           &
          &
          -\hat\varphi_{-1}^\dagger(\bm{x})    \hat\varphi_{0}(\bm{x}') 
          &
          \hat\varphi_{-1}^\dagger(\bm{x})    \hat\varphi_{1}(\bm{x}') 
          &
          -\hat\varphi_{-1}^\dagger(\bm{x})    \hat\varphi_{2}(\bm{x}') 
\\
        \hat\varphi_{0}^\dagger(\bm{x})    \hat\varphi_{-2}(\bm{x}') 
          &
          -\hat\varphi_{0}^\dagger(\bm{x})    \hat\varphi_{-1}(\bm{x}')           &
          &
          \hat\varphi_{0}^\dagger(\bm{x})    \hat\varphi_{0}(\bm{x}') 
          &
         - \hat\varphi_{0}^\dagger(\bm{x})    \hat\varphi_{1}(\bm{x}') 
          &
          \hat\varphi_{0}^\dagger(\bm{x})    \hat\varphi_{2}(\bm{x}') 
\\
        -\hat\varphi_{1}^\dagger(\bm{x})    \hat\varphi_{-2}(\bm{x}') 
          &
          \hat\varphi_{1}^\dagger(\bm{x})    \hat\varphi_{-1}(\bm{x}')           &
          &
         - \hat\varphi_{1}^\dagger(\bm{x})    \hat\varphi_{0}(\bm{x}') 
          &
          \hat\varphi_{1}^\dagger(\bm{x})    \hat\varphi_{1}(\bm{x}') 
          &
          -\hat\varphi_{1}^\dagger(\bm{x})    \hat\varphi_{2}(\bm{x}') 
\\
          \hat\varphi_{2}^\dagger(\bm{x})    \hat\varphi_{-2}(\bm{x}') 
          &
          -\hat\varphi_{2}^\dagger(\bm{x})    \hat\varphi_{-1}(\bm{x}')           &
          &
          \hat\varphi_{2}^\dagger(\bm{x})    \hat\varphi_{0}(\bm{x}') 
          &
          -\hat\varphi_{2}^\dagger(\bm{x})    \hat\varphi_{1}(\bm{x}') 
          &
          \hat\varphi_{2}^\dagger(\bm{x})    \hat\varphi_{2}(\bm{x}') 
\end{pmatrix}  .
\end{align}
\end{subequations}
\end{widetext}
 With the 2-point number density $\rho(\bm{x},\bm{x}') =  \hat{\varphi}^\dagger(\bm{x}) \hat{\varphi}(\bm{x}')$, it is possible to define the dimensionless 2-point number generator
\begin{align}
\label{A00_two_point_N_operator}
   \hat{\cal N}(\bm{x},\bm{x}') 
   &\equiv
   \frac{ \hat N^{00}(\bm{x},\bm{x}')}{\rho(\bm{x},\bm{x}')/5}
   ,
\end{align}
which is an  2-point idempotent operator
\begin{align}
   \hat{\cal N}(\bm{x},\bm{x}')^2 
   &=
   \hat{\cal N}(\bm{x},\bm{x}')
   .
\end{align}
Therefore, the equation of motion  may be written as
\begin{subequations}
\label{equation_of_motion_spin_2_BEC_point_contact}
\begin{align}
\nonumber
i\hbar \partial_t \hat{\varphi}
&\stackrel{(\ref{equation_of_motion_spinor_BEC})}{=}
  \left(
 -\frac{\hbar^2}{2m}\nabla^2 - \mu
 +
  g_0 \hat{\varphi}^\dagger \hat{\varphi}
 \right)\hat{\varphi}
 \\
 & +
 \frac{ g_1}{\hbar^2}  {\hat{\bm{F}}}. \bm{f}\hat{\varphi}
 +
\frac{ g_2}{2}  
  \hat N^{00}
 \hat{\varphi}
 \\
 \nonumber
&\stackrel{(\ref{A00_two_point_N_operator})}{=}
  \left(
 -\frac{\hbar^2}{2m}\nabla^2 - \mu
 +
  g_0 \hat{\varphi}^\dagger \hat{\varphi}
 \right)\hat{\varphi}
\\
& +
 \frac{ g_1}{\hbar^2}  {\hat{\bm{F}}}. \bm{f}\hat{\varphi}
  +
\frac{ g_2 \rho}{10}  
 \hat{\cal N}
 \hat{\varphi}
 .
\end{align}
\end{subequations}
Therefore, $\exp\left[ 
 -i \left(
 {g_2 \hat N^{00} \otimes \bm{1}}{2}
  \right)
  {\tau}/{\hbar}
   \right]$ in $U^\text{\tiny nd}_{f=2} [\hat{\varphi}]$ on the righthand side of (\ref{split_unitary_operators})  can  be analytically expanded to all orders
\begin{subequations}
\label{nondiagonal_N00_equation_of_motion_spin_2_unitary_form_analytical_expansion}
\begin{align}
U^\text{\tiny nd}_{f=2}  [\hat A_{00},\hat{\varphi}]
& =
 1
 +
 \left(
 \exp\left(-\frac{i g_2 \rho\tau}{10\hbar}\right) 
-1\right) 
{{\cal N}}
.
\label{nondiagonal_N00_equation_of_motion_spin_2_unitary_form_analytical_expansion_b}
\end{align}
\end{subequations}
 Finally, a quantum algorithm (in split unitary operator form applicable to the spin-2 BEC) with strong spin-density coupling (even for $g_1\sim 1$)  and $\hat N^{00}$ coupling (even for $g_2\sim 1$)   is
\begin{align}
\label{spin_2_BEC_quantum_algorithm}
\hat{\psi}(\bm{x}, t+\tau)  
\stackrel{(\ref{split_unitary_operators})}{=}
U^\text{\tiny nd}_{f=2}  [\hat{\psi}]
U^\text{\tiny d} [\hat{\psi}] 
\,
\hat{\psi}(\bm{x},t),
\end{align}
where 
\begin{widetext}
\begin{align}
   U^\text{\tiny nd}_{f=2}  [\hat{\psi}]
   & \stackrel{(\ref{nondiagonal_N00_equation_of_motion_spin_2_unitary_form_analytical_expansion})}{\stackrel{(\ref{nondiagonal_spin_density_evolution_operator})}{=}}
\left(U^\text{\tiny nd}_{f=1/2}  [{\hat{\bm{F}}}_{({1}/{2})},\hat{\varphi}]
U^\text{\tiny nd}_{f=1}  [{\hat{\bm{F}}}_{(1)} ,\hat{\varphi}]
U^\text{\tiny nd}_{f=2}  [\hat A_{00},\hat{\varphi}]
\right) \otimes \bm{1}.
 \end{align}
This completes the description of the quantum lattice gas representation of non-Abelian spinor BECs. 
\vspace{1em}

\end{widetext}

\section{Conclusion}

  A closed-form analytical expansion of the nondiagonal part (spin density coupling) of the hyperfine interaction of a non-Abelian superfluid (zero-temperature spinor-2 BEC) was presented. Spin-2 BEC theory possesses non-Abelian  gauge symmetry, so spin-2 quantum matter in its  strong coupling regime is not amenable to perturbative treatment.
  Yet, it is possible to analytically continue the representation into the nonperturbative (strong-coupling) regime $g_1\sim 1$ by matching the nondiagonal collide operators in the perturbative (weak-coupling)  regime $g_1\lll 1$.

In the many-body sector (e.g. for spin-2 BECs containing particle-particle entangled states) (\ref{spin_2_BEC_quantum_algorithm})  can also serve as an efficient quantum algorithm for future implementation on a Feynman quantum computer \cite{feynman-ces60,feynman-82,feynman-85}. 
  In principle on a Feynman quantum computer, such quantum algorithms can be a more precise tool for studying spinor BEC dynamics. It can exceed the utility of analog quantum simulators based on ultracold quantum gases of alkali atoms  because of the absence of experimental noise and related decoherence effects in spinor BEC experiments relying on engineered Hamiltonians. 

  In the single-body sector and  in the low-energy limit, the quantum algorithm on the analytically-continued spin-density interaction (\ref{spin_2_BEC_quantum_algorithm}) is congruent to spin-2 GP equations, which is a representation of the spin-2 BEC at zero temperature, and can serve as an accurate unitary algorithm  for  supercomputer implementation today. 
  Numerical quantum simulations of the $f=2$ spinor BEC  based on quantum algorithm (\ref{spin_2_BEC_quantum_algorithm})  applied to the time-dependent interaction of non-Abelian quantum vortices will be presented in a subsequent communication.

\section{Acknowledgements}

I would like to thank Prof. George Vahala for helpful discussions about spinor BECs. I would like to thank my graduate students Jasper Taylor and Steven Smith for their help in implement this quantum algorithm for a spin-2 BEC and the calculating the L2 norm numerical accuracy.  This work was supported by the grant  ``Quantum Computational Mathematics for Efficient Computational Physics"  from the Air Force Office of Scientific Research. 


\begin{thebibliography}{39}
\expandafter\ifx\csname natexlab\endcsname\relax\def\natexlab#1{#1}\fi
\expandafter\ifx\csname bibnamefont\endcsname\relax
  \def\bibnamefont#1{#1}\fi
\expandafter\ifx\csname bibfnamefont\endcsname\relax
  \def\bibfnamefont#1{#1}\fi
\expandafter\ifx\csname citenamefont\endcsname\relax
  \def\citenamefont#1{#1}\fi
\expandafter\ifx\csname url\endcsname\relax
  \def\url#1{\texttt{#1}}\fi
\expandafter\ifx\csname urlprefix\endcsname\relax\def\urlprefix{URL }\fi
\providecommand{\bibinfo}[2]{#2}
\providecommand{\eprint}[2][]{\url{#2}}

\bibitem[{\citenamefont{Ketterle}(2002)}]{RevModPhys.74.1131}
\bibinfo{author}{\bibfnamefont{W.}~\bibnamefont{Ketterle}},
  \bibinfo{journal}{Rev. Mod. Phys.} \textbf{\bibinfo{volume}{74}},
  \bibinfo{pages}{1131} (\bibinfo{year}{2002}).

\bibitem[{\citenamefont{Einstein}(1925)}]{1925SPAW1_3}
\bibinfo{author}{\bibfnamefont{A.}~\bibnamefont{Einstein}},
  \bibinfo{journal}{Sitzungsberichte der Preussischen Akademie der
  Wissenschaften zu Berlin} \textbf{\bibinfo{volume}{1}}, \bibinfo{pages}{261}
  (\bibinfo{year}{1925}).

\bibitem[{\citenamefont{Anderson et~al.}(1995)\citenamefont{Anderson, Ensher,
  Matthews, Wieman, and Cornell}}]{Anderson:1995p2538}
\bibinfo{author}{\bibfnamefont{M.}~\bibnamefont{Anderson}},
  \bibinfo{author}{\bibfnamefont{J.}~\bibnamefont{Ensher}},
  \bibinfo{author}{\bibfnamefont{M.}~\bibnamefont{Matthews}},
  \bibinfo{author}{\bibfnamefont{C.}~\bibnamefont{Wieman}}, \bibnamefont{and}
  \bibinfo{author}{\bibfnamefont{E.}~\bibnamefont{Cornell}},
  \bibinfo{journal}{Science} \textbf{\bibinfo{volume}{269}},
  \bibinfo{pages}{198} (\bibinfo{year}{1995}).

\bibitem[{\citenamefont{Davis et~al.}(1995)\citenamefont{Davis, Mewes, Andrews,
  van Druten, Durfee, Kurn, and Ketterle}}]{PhysRevLett.75.3969}
\bibinfo{author}{\bibfnamefont{K.~B.} \bibnamefont{Davis}},
  \bibinfo{author}{\bibfnamefont{M.~O.} \bibnamefont{Mewes}},
  \bibinfo{author}{\bibfnamefont{M.~R.} \bibnamefont{Andrews}},
  \bibinfo{author}{\bibfnamefont{N.~J.} \bibnamefont{van Druten}},
  \bibinfo{author}{\bibfnamefont{D.~S.} \bibnamefont{Durfee}},
  \bibinfo{author}{\bibfnamefont{D.~M.} \bibnamefont{Kurn}}, \bibnamefont{and}
  \bibinfo{author}{\bibfnamefont{W.}~\bibnamefont{Ketterle}},
  \bibinfo{journal}{Phys. Rev. Lett.} \textbf{\bibinfo{volume}{75}},
  \bibinfo{pages}{3969} (\bibinfo{year}{1995}).

\bibitem[{\citenamefont{Mewes et~al.}(1996)\citenamefont{Mewes, Andrews, van
  Druten, Kurn, Durfee, and Ketterle}}]{PhysRevLett.77.416}
\bibinfo{author}{\bibfnamefont{M.-O.} \bibnamefont{Mewes}},
  \bibinfo{author}{\bibfnamefont{M.~R.} \bibnamefont{Andrews}},
  \bibinfo{author}{\bibfnamefont{N.~J.} \bibnamefont{van Druten}},
  \bibinfo{author}{\bibfnamefont{D.~M.} \bibnamefont{Kurn}},
  \bibinfo{author}{\bibfnamefont{D.~S.} \bibnamefont{Durfee}},
  \bibnamefont{and} \bibinfo{author}{\bibfnamefont{W.}~\bibnamefont{Ketterle}},
  \bibinfo{journal}{Phys. Rev. Lett.} \textbf{\bibinfo{volume}{77}},
  \bibinfo{pages}{416} (\bibinfo{year}{1996}).

\bibitem[{\citenamefont{Du et~al.}(2004)\citenamefont{Du, Squires, Imai, Czaia,
  Saravanan, Bright, Reichel, H\"ansch, and Anderson}}]{PhysRevA.70.053606}
\bibinfo{author}{\bibfnamefont{S.}~\bibnamefont{Du}},
  \bibinfo{author}{\bibfnamefont{M.~B.} \bibnamefont{Squires}},
  \bibinfo{author}{\bibfnamefont{Y.}~\bibnamefont{Imai}},
  \bibinfo{author}{\bibfnamefont{L.}~\bibnamefont{Czaia}},
  \bibinfo{author}{\bibfnamefont{R.~A.} \bibnamefont{Saravanan}},
  \bibinfo{author}{\bibfnamefont{V.}~\bibnamefont{Bright}},
  \bibinfo{author}{\bibfnamefont{J.}~\bibnamefont{Reichel}},
  \bibinfo{author}{\bibfnamefont{T.~W.} \bibnamefont{H\"ansch}},
  \bibnamefont{and} \bibinfo{author}{\bibfnamefont{D.~Z.}
  \bibnamefont{Anderson}}, \bibinfo{journal}{Phys. Rev. A}
  \textbf{\bibinfo{volume}{70}}, \bibinfo{pages}{053606}
  (\bibinfo{year}{2004}),
  \urlprefix\url{http://link.aps.org/doi/10.1103/PhysRevA.70.053606}.

\bibitem[{\citenamefont{Moritz et~al.}(2003)\citenamefont{Moritz, St\"oferle,
  K\"ohl, and Esslinger}}]{PhysRevLett.91.250402}
\bibinfo{author}{\bibfnamefont{H.}~\bibnamefont{Moritz}},
  \bibinfo{author}{\bibfnamefont{T.}~\bibnamefont{St\"oferle}},
  \bibinfo{author}{\bibfnamefont{M.}~\bibnamefont{K\"ohl}}, \bibnamefont{and}
  \bibinfo{author}{\bibfnamefont{T.}~\bibnamefont{Esslinger}},
  \bibinfo{journal}{Phys. Rev. Lett.} \textbf{\bibinfo{volume}{91}},
  \bibinfo{pages}{250402} (\bibinfo{year}{2003}),
  \urlprefix\url{http://link.aps.org/doi/10.1103/PhysRevLett.91.250402}.

\bibitem[{\citenamefont{St\"oferle et~al.}(2004)\citenamefont{St\"oferle,
  Moritz, Schori, K\"ohl, and Esslinger}}]{PhysRevLett.92.130403}
\bibinfo{author}{\bibfnamefont{T.}~\bibnamefont{St\"oferle}},
  \bibinfo{author}{\bibfnamefont{H.}~\bibnamefont{Moritz}},
  \bibinfo{author}{\bibfnamefont{C.}~\bibnamefont{Schori}},
  \bibinfo{author}{\bibfnamefont{M.}~\bibnamefont{K\"ohl}}, \bibnamefont{and}
  \bibinfo{author}{\bibfnamefont{T.}~\bibnamefont{Esslinger}},
  \bibinfo{journal}{Phys. Rev. Lett.} \textbf{\bibinfo{volume}{92}},
  \bibinfo{pages}{130403} (\bibinfo{year}{2004}),
  \urlprefix\url{http://link.aps.org/doi/10.1103/PhysRevLett.92.130403}.

\bibitem[{\citenamefont{Miyake et~al.}(2013)\citenamefont{Miyake, Siviloglou,
  Kennedy, Burton, and Ketterle}}]{PhysRevLett.111.185302}
\bibinfo{author}{\bibfnamefont{H.}~\bibnamefont{Miyake}},
  \bibinfo{author}{\bibfnamefont{G.~A.} \bibnamefont{Siviloglou}},
  \bibinfo{author}{\bibfnamefont{C.~J.} \bibnamefont{Kennedy}},
  \bibinfo{author}{\bibfnamefont{W.~C.} \bibnamefont{Burton}},
  \bibnamefont{and} \bibinfo{author}{\bibfnamefont{W.}~\bibnamefont{Ketterle}},
  \bibinfo{journal}{Phys. Rev. Lett.} \textbf{\bibinfo{volume}{111}},
  \bibinfo{pages}{185302} (\bibinfo{year}{2013}),
  \urlprefix\url{http://link.aps.org/doi/10.1103/PhysRevLett.111.185302}.

\bibitem[{\citenamefont{Aidelsburger et~al.}(2011)\citenamefont{Aidelsburger,
  Atala, Nascimb\`ene, Trotzky, Chen, and Bloch}}]{PhysRevLett.107.255301}
\bibinfo{author}{\bibfnamefont{M.}~\bibnamefont{Aidelsburger}},
  \bibinfo{author}{\bibfnamefont{M.}~\bibnamefont{Atala}},
  \bibinfo{author}{\bibfnamefont{S.}~\bibnamefont{Nascimb\`ene}},
  \bibinfo{author}{\bibfnamefont{S.}~\bibnamefont{Trotzky}},
  \bibinfo{author}{\bibfnamefont{Y.-A.} \bibnamefont{Chen}}, \bibnamefont{and}
  \bibinfo{author}{\bibfnamefont{I.}~\bibnamefont{Bloch}},
  \bibinfo{journal}{Phys. Rev. Lett.} \textbf{\bibinfo{volume}{107}},
  \bibinfo{pages}{255301} (\bibinfo{year}{2011}),
  \urlprefix\url{http://link.aps.org/doi/10.1103/PhysRevLett.107.255301}.

\bibitem[{\citenamefont{Tagliacozzo
  et~al.}(2013{\natexlab{a}})\citenamefont{Tagliacozzo, Celi, Orland, Mitchell,
  and Lewenstein}}]{ncomms3615}
\bibinfo{author}{\bibfnamefont{L.}~\bibnamefont{Tagliacozzo}},
  \bibinfo{author}{\bibfnamefont{A.}~\bibnamefont{Celi}},
  \bibinfo{author}{\bibfnamefont{P.}~\bibnamefont{Orland}},
  \bibinfo{author}{\bibfnamefont{M.~W.} \bibnamefont{Mitchell}},
  \bibnamefont{and}
  \bibinfo{author}{\bibfnamefont{M.}~\bibnamefont{Lewenstein}},
  \bibinfo{journal}{Nature Communications} \textbf{\bibinfo{volume}{4}}
  (\bibinfo{year}{2013}{\natexlab{a}}),
  \urlprefix\url{http://dx.doi.org/10.1038/ncomms3615}.

\bibitem[{\citenamefont{Farkas et~al.}(2013)\citenamefont{Farkas, Hudek, Du,
  and Anderson}}]{PhysRevA.87.053417}
\bibinfo{author}{\bibfnamefont{D.~M.} \bibnamefont{Farkas}},
  \bibinfo{author}{\bibfnamefont{K.~M.} \bibnamefont{Hudek}},
  \bibinfo{author}{\bibfnamefont{S.}~\bibnamefont{Du}}, \bibnamefont{and}
  \bibinfo{author}{\bibfnamefont{D.~Z.} \bibnamefont{Anderson}},
  \bibinfo{journal}{Phys. Rev. A} \textbf{\bibinfo{volume}{87}},
  \bibinfo{pages}{053417} (\bibinfo{year}{2013}),
  \urlprefix\url{http://link.aps.org/doi/10.1103/PhysRevA.87.053417}.

\bibitem[{\citenamefont{Ciobanu et~al.}(2000)\citenamefont{Ciobanu, Yip, and
  Ho}}]{PhysRevA.61.033607}
\bibinfo{author}{\bibfnamefont{C.~V.} \bibnamefont{Ciobanu}},
  \bibinfo{author}{\bibfnamefont{S.-K.} \bibnamefont{Yip}}, \bibnamefont{and}
  \bibinfo{author}{\bibfnamefont{T.-L.} \bibnamefont{Ho}},
  \bibinfo{journal}{Phys. Rev. A} \textbf{\bibinfo{volume}{61}},
  \bibinfo{pages}{033607} (\bibinfo{year}{2000}),
  \urlprefix\url{http://link.aps.org/doi/10.1103/PhysRevA.61.033607}.

\bibitem[{\citenamefont{Zheng and Brun}(2012)}]{PhysRevA.86.032323}
\bibinfo{author}{\bibfnamefont{Y.-C.} \bibnamefont{Zheng}} \bibnamefont{and}
  \bibinfo{author}{\bibfnamefont{T.~A.} \bibnamefont{Brun}},
  \bibinfo{journal}{Phys. Rev. A} \textbf{\bibinfo{volume}{86}},
  \bibinfo{pages}{032323} (\bibinfo{year}{2012}),
  \urlprefix\url{http://link.aps.org/doi/10.1103/PhysRevA.86.032323}.

\bibitem[{\citenamefont{Garay et~al.}(2000)\citenamefont{Garay, Anglin, Cirac,
  and Zoller}}]{PhysRevLett.85.4643}
\bibinfo{author}{\bibfnamefont{L.~J.} \bibnamefont{Garay}},
  \bibinfo{author}{\bibfnamefont{J.~R.} \bibnamefont{Anglin}},
  \bibinfo{author}{\bibfnamefont{J.~I.} \bibnamefont{Cirac}}, \bibnamefont{and}
  \bibinfo{author}{\bibfnamefont{P.}~\bibnamefont{Zoller}},
  \bibinfo{journal}{Phys. Rev. Lett.} \textbf{\bibinfo{volume}{85}},
  \bibinfo{pages}{4643} (\bibinfo{year}{2000}),
  \urlprefix\url{http://link.aps.org/doi/10.1103/PhysRevLett.85.4643}.

\bibitem[{\citenamefont{Zohar and Reznik}(2011)}]{PhysRevLett.107.275301}
\bibinfo{author}{\bibfnamefont{E.}~\bibnamefont{Zohar}} \bibnamefont{and}
  \bibinfo{author}{\bibfnamefont{B.}~\bibnamefont{Reznik}},
  \bibinfo{journal}{Phys. Rev. Lett.} \textbf{\bibinfo{volume}{107}},
  \bibinfo{pages}{275301} (\bibinfo{year}{2011}),
  \urlprefix\url{http://link.aps.org/doi/10.1103/PhysRevLett.107.275301}.

\bibitem[{\citenamefont{Cirac et~al.}(2010)\citenamefont{Cirac, Maraner, and
  Pachos}}]{PhysRevLett.105.190403}
\bibinfo{author}{\bibfnamefont{J.~I.} \bibnamefont{Cirac}},
  \bibinfo{author}{\bibfnamefont{P.}~\bibnamefont{Maraner}}, \bibnamefont{and}
  \bibinfo{author}{\bibfnamefont{J.~K.} \bibnamefont{Pachos}},
  \bibinfo{journal}{Phys. Rev. Lett.} \textbf{\bibinfo{volume}{105}},
  \bibinfo{pages}{190403} (\bibinfo{year}{2010}),
  \urlprefix\url{http://link.aps.org/doi/10.1103/PhysRevLett.105.190403}.

\bibitem[{\citenamefont{Tagliacozzo
  et~al.}(2013{\natexlab{b}})\citenamefont{Tagliacozzo, Celi, Zamora, and
  Lewenstein}}]{Tagliacozzo2013160}
\bibinfo{author}{\bibfnamefont{L.}~\bibnamefont{Tagliacozzo}},
  \bibinfo{author}{\bibfnamefont{A.}~\bibnamefont{Celi}},
  \bibinfo{author}{\bibfnamefont{A.}~\bibnamefont{Zamora}}, \bibnamefont{and}
  \bibinfo{author}{\bibfnamefont{M.}~\bibnamefont{Lewenstein}},
  \bibinfo{journal}{Annals of Physics} \textbf{\bibinfo{volume}{330}},
  \bibinfo{pages}{160 } (\bibinfo{year}{2013}{\natexlab{b}}), ISSN
  \bibinfo{issn}{0003-4916},
  \urlprefix\url{http://www.sciencedirect.com/science/article/pii/S0003491612001819}.

\bibitem[{\citenamefont{Khawaja and Stoof}(2001)}]{PhysRevA.64.043612}
\bibinfo{author}{\bibfnamefont{U.~A.} \bibnamefont{Khawaja}} \bibnamefont{and}
  \bibinfo{author}{\bibfnamefont{H.~T.~C.} \bibnamefont{Stoof}},
  \bibinfo{journal}{Phys. Rev. A} \textbf{\bibinfo{volume}{64}},
  \bibinfo{pages}{043612} (\bibinfo{year}{2001}),
  \urlprefix\url{http://link.aps.org/doi/10.1103/PhysRevA.64.043612}.

\bibitem[{\citenamefont{Wang et~al.}(2005)\citenamefont{Wang, Anderson, Bright,
  Cornell, Diot, Kishimoto, Prentiss, Saravanan, Segal, and
  Wu}}]{PhysRevLett.94.090405}
\bibinfo{author}{\bibfnamefont{Y.-J.} \bibnamefont{Wang}},
  \bibinfo{author}{\bibfnamefont{D.~Z.} \bibnamefont{Anderson}},
  \bibinfo{author}{\bibfnamefont{V.~M.} \bibnamefont{Bright}},
  \bibinfo{author}{\bibfnamefont{E.~A.} \bibnamefont{Cornell}},
  \bibinfo{author}{\bibfnamefont{Q.}~\bibnamefont{Diot}},
  \bibinfo{author}{\bibfnamefont{T.}~\bibnamefont{Kishimoto}},
  \bibinfo{author}{\bibfnamefont{M.}~\bibnamefont{Prentiss}},
  \bibinfo{author}{\bibfnamefont{R.~A.} \bibnamefont{Saravanan}},
  \bibinfo{author}{\bibfnamefont{S.~R.} \bibnamefont{Segal}}, \bibnamefont{and}
  \bibinfo{author}{\bibfnamefont{S.}~\bibnamefont{Wu}}, \bibinfo{journal}{Phys.
  Rev. Lett.} \textbf{\bibinfo{volume}{94}}, \bibinfo{pages}{090405}
  (\bibinfo{year}{2005}),
  \urlprefix\url{http://link.aps.org/doi/10.1103/PhysRevLett.94.090405}.

\bibitem[{\citenamefont{Shin et~al.}(2005)\citenamefont{Shin, Sanner, Jo,
  Pasquini, Saba, Ketterle, Pritchard, Vengalattore, and
  Prentiss}}]{PhysRevA.72.021604}
\bibinfo{author}{\bibfnamefont{Y.}~\bibnamefont{Shin}},
  \bibinfo{author}{\bibfnamefont{C.}~\bibnamefont{Sanner}},
  \bibinfo{author}{\bibfnamefont{G.-B.} \bibnamefont{Jo}},
  \bibinfo{author}{\bibfnamefont{T.~A.} \bibnamefont{Pasquini}},
  \bibinfo{author}{\bibfnamefont{M.}~\bibnamefont{Saba}},
  \bibinfo{author}{\bibfnamefont{W.}~\bibnamefont{Ketterle}},
  \bibinfo{author}{\bibfnamefont{D.~E.} \bibnamefont{Pritchard}},
  \bibinfo{author}{\bibfnamefont{M.}~\bibnamefont{Vengalattore}},
  \bibnamefont{and} \bibinfo{author}{\bibfnamefont{M.}~\bibnamefont{Prentiss}},
  \bibinfo{journal}{Phys. Rev. A} \textbf{\bibinfo{volume}{72}},
  \bibinfo{pages}{021604} (\bibinfo{year}{2005}),
  \urlprefix\url{http://link.aps.org/doi/10.1103/PhysRevA.72.021604}.

\bibitem[{\citenamefont{Jo et~al.}(2007{\natexlab{a}})\citenamefont{Jo, Shin,
  Will, Pasquini, Saba, Ketterle, Pritchard, Vengalattore, and
  Prentiss}}]{PhysRevLett.98.030407}
\bibinfo{author}{\bibfnamefont{G.-B.} \bibnamefont{Jo}},
  \bibinfo{author}{\bibfnamefont{Y.}~\bibnamefont{Shin}},
  \bibinfo{author}{\bibfnamefont{S.}~\bibnamefont{Will}},
  \bibinfo{author}{\bibfnamefont{T.~A.} \bibnamefont{Pasquini}},
  \bibinfo{author}{\bibfnamefont{M.}~\bibnamefont{Saba}},
  \bibinfo{author}{\bibfnamefont{W.}~\bibnamefont{Ketterle}},
  \bibinfo{author}{\bibfnamefont{D.~E.} \bibnamefont{Pritchard}},
  \bibinfo{author}{\bibfnamefont{M.}~\bibnamefont{Vengalattore}},
  \bibnamefont{and} \bibinfo{author}{\bibfnamefont{M.}~\bibnamefont{Prentiss}},
  \bibinfo{journal}{Phys. Rev. Lett.} \textbf{\bibinfo{volume}{98}},
  \bibinfo{pages}{030407} (\bibinfo{year}{2007}{\natexlab{a}}),
  \urlprefix\url{http://link.aps.org/doi/10.1103/PhysRevLett.98.030407}.

\bibitem[{\citenamefont{Jo et~al.}(2007{\natexlab{b}})\citenamefont{Jo, Choi,
  Christensen, Pasquini, Lee, Ketterle, and Pritchard}}]{PhysRevLett.98.180401}
\bibinfo{author}{\bibfnamefont{G.-B.} \bibnamefont{Jo}},
  \bibinfo{author}{\bibfnamefont{J.-H.} \bibnamefont{Choi}},
  \bibinfo{author}{\bibfnamefont{C.~A.} \bibnamefont{Christensen}},
  \bibinfo{author}{\bibfnamefont{T.~A.} \bibnamefont{Pasquini}},
  \bibinfo{author}{\bibfnamefont{Y.-R.} \bibnamefont{Lee}},
  \bibinfo{author}{\bibfnamefont{W.}~\bibnamefont{Ketterle}}, \bibnamefont{and}
  \bibinfo{author}{\bibfnamefont{D.~E.} \bibnamefont{Pritchard}},
  \bibinfo{journal}{Phys. Rev. Lett.} \textbf{\bibinfo{volume}{98}},
  \bibinfo{pages}{180401} (\bibinfo{year}{2007}{\natexlab{b}}),
  \urlprefix\url{http://link.aps.org/doi/10.1103/PhysRevLett.98.180401}.

\bibitem[{\citenamefont{Stickney et~al.}(2007)\citenamefont{Stickney, Anderson,
  and Zozulya}}]{PhysRevA.75.063603}
\bibinfo{author}{\bibfnamefont{J.~A.} \bibnamefont{Stickney}},
  \bibinfo{author}{\bibfnamefont{D.~Z.} \bibnamefont{Anderson}},
  \bibnamefont{and} \bibinfo{author}{\bibfnamefont{A.~A.}
  \bibnamefont{Zozulya}}, \bibinfo{journal}{Phys. Rev. A}
  \textbf{\bibinfo{volume}{75}}, \bibinfo{pages}{063603}
  (\bibinfo{year}{2007}),
  \urlprefix\url{http://link.aps.org/doi/10.1103/PhysRevA.75.063603}.

\bibitem[{\citenamefont{Stickney et~al.}(2008)\citenamefont{Stickney, Kafle,
  Anderson, and Zozulya}}]{PhysRevA.77.043604}
\bibinfo{author}{\bibfnamefont{J.~A.} \bibnamefont{Stickney}},
  \bibinfo{author}{\bibfnamefont{R.~P.} \bibnamefont{Kafle}},
  \bibinfo{author}{\bibfnamefont{D.~Z.} \bibnamefont{Anderson}},
  \bibnamefont{and} \bibinfo{author}{\bibfnamefont{A.~A.}
  \bibnamefont{Zozulya}}, \bibinfo{journal}{Phys. Rev. A}
  \textbf{\bibinfo{volume}{77}}, \bibinfo{pages}{043604}
  (\bibinfo{year}{2008}),
  \urlprefix\url{http://link.aps.org/doi/10.1103/PhysRevA.77.043604}.

\bibitem[{\citenamefont{Carter et~al.}(2012)\citenamefont{Carter, Cherry, and
  Martin}}]{PhysRevA.86.053401}
\bibinfo{author}{\bibfnamefont{J.~D.} \bibnamefont{Carter}},
  \bibinfo{author}{\bibfnamefont{O.}~\bibnamefont{Cherry}}, \bibnamefont{and}
  \bibinfo{author}{\bibfnamefont{J.~D.~D.} \bibnamefont{Martin}},
  \bibinfo{journal}{Phys. Rev. A} \textbf{\bibinfo{volume}{86}},
  \bibinfo{pages}{053401} (\bibinfo{year}{2012}),
  \urlprefix\url{http://link.aps.org/doi/10.1103/PhysRevA.86.053401}.

\bibitem[{\citenamefont{Trotter}(1959)}]{Trotter_JSTOR_1959}
\bibinfo{author}{\bibfnamefont{H.~F.} \bibnamefont{Trotter}},
  \bibinfo{journal}{Proceedings of the American Mathematical Society}
  \textbf{\bibinfo{volume}{10}}, \bibinfo{pages}{pp. 545}
  (\bibinfo{year}{1959}), ISSN \bibinfo{issn}{00029939},
  \urlprefix\url{http://www.jstor.org/stable/2033649}.

\bibitem[{\citenamefont{Ueda and Kawaguchi}(2010)}]{ueda_10}
\bibinfo{author}{\bibfnamefont{M.}~\bibnamefont{Ueda}} \bibnamefont{and}
  \bibinfo{author}{\bibfnamefont{Y.}~\bibnamefont{Kawaguchi}},
  \bibinfo{journal}{arXiv:1001.2072 [cond-mat.quant-gas]}
  \bibinfo{eid}{arXiv:cond-mat.quant-gas.1001.2072v2} (pages
  \bibinfo{numpages}{123}) (\bibinfo{year}{2010}),
  \urlprefix\url{http://arxiv.org/abs/1001.2072}.

\bibitem[{\citenamefont{Ando et~al.}(2011)\citenamefont{Ando, Ohtake, Kondo,
  and Nakamura}}]{PhysRevA.83.023619}
\bibinfo{author}{\bibfnamefont{T.}~\bibnamefont{Ando}},
  \bibinfo{author}{\bibfnamefont{Y.}~\bibnamefont{Ohtake}},
  \bibinfo{author}{\bibfnamefont{J.-i.} \bibnamefont{Kondo}}, \bibnamefont{and}
  \bibinfo{author}{\bibfnamefont{K.}~\bibnamefont{Nakamura}},
  \bibinfo{journal}{Phys. Rev. A} \textbf{\bibinfo{volume}{83}},
  \bibinfo{pages}{023619} (\bibinfo{year}{2011}),
  \urlprefix\url{http://link.aps.org/doi/10.1103/PhysRevA.83.023619}.

\bibitem[{\citenamefont{Yepez and Boghosian}(2002)}]{yepez-cpc01}
\bibinfo{author}{\bibfnamefont{J.}~\bibnamefont{Yepez}} \bibnamefont{and}
  \bibinfo{author}{\bibfnamefont{B.}~\bibnamefont{Boghosian}},
  \bibinfo{journal}{Computer Physics Communications}
  \textbf{\bibinfo{volume}{146}}, \bibinfo{pages}{280} (\bibinfo{year}{2002}).

\bibitem[{\citenamefont{Yepez et~al.}(2009{\natexlab{a}})\citenamefont{Yepez,
  Vahala, and Vahala}}]{yepez-vahala-EPJ-09}
\bibinfo{author}{\bibfnamefont{J.}~\bibnamefont{Yepez}},
  \bibinfo{author}{\bibfnamefont{G.}~\bibnamefont{Vahala}}, \bibnamefont{and}
  \bibinfo{author}{\bibfnamefont{L.}~\bibnamefont{Vahala}},
  \bibinfo{journal}{Euro. Phys. J. Special Topics}
  \textbf{\bibinfo{volume}{171}}, \bibinfo{pages}{9}
  (\bibinfo{year}{2009}{\natexlab{a}}).

\bibitem[{\citenamefont{Yepez et~al.}(2009{\natexlab{b}})\citenamefont{Yepez,
  Vahala, Vahala, and Soe}}]{yepez:084501}
\bibinfo{author}{\bibfnamefont{J.}~\bibnamefont{Yepez}},
  \bibinfo{author}{\bibfnamefont{G.}~\bibnamefont{Vahala}},
  \bibinfo{author}{\bibfnamefont{L.}~\bibnamefont{Vahala}}, \bibnamefont{and}
  \bibinfo{author}{\bibfnamefont{M.}~\bibnamefont{Soe}},
  \bibinfo{journal}{Physical Review Letters} \textbf{\bibinfo{volume}{103}},
  \bibinfo{eid}{084501} (pages~\bibinfo{numpages}{4})
  (\bibinfo{year}{2009}{\natexlab{b}}),
  \urlprefix\url{http://link.aps.org/abstract/PRL/v103/e084501}.

\bibitem[{\citenamefont{Yepez et~al.}(2010)\citenamefont{Yepez, Vahala, Vahala,
  and Soe}}]{yepez:770209}
\bibinfo{author}{\bibfnamefont{J.}~\bibnamefont{Yepez}},
  \bibinfo{author}{\bibfnamefont{G.}~\bibnamefont{Vahala}},
  \bibinfo{author}{\bibfnamefont{L.}~\bibnamefont{Vahala}}, \bibnamefont{and}
  \bibinfo{author}{\bibfnamefont{M.}~\bibnamefont{Soe}}, in
  \emph{\bibinfo{booktitle}{Quantum Information and Computation VIII}}, edited
  by \bibinfo{editor}{\bibfnamefont{E.~J.} \bibnamefont{Donkor}},
  \bibinfo{editor}{\bibfnamefont{A.~R.} \bibnamefont{Pirich}},
  \bibnamefont{and} \bibinfo{editor}{\bibfnamefont{H.~E.} \bibnamefont{Brandt}}
  (\bibinfo{publisher}{SPIE}, \bibinfo{year}{2010}), vol.
  \bibinfo{volume}{7702}, p. \bibinfo{pages}{770209},
  \urlprefix\url{http://link.aip.org/link/?PSI/7702/770209/1}.

\bibitem[{\citenamefont{Yepez}(2016)}]{yepez_arXiv1609.02229_cond_mat.quant_gas}
\bibinfo{author}{\bibfnamefont{J.}~\bibnamefont{Yepez}},
  \bibinfo{journal}{arXiv:1609.02229 [cond-mat.quant-gas]}
  (\bibinfo{year}{2016}).

\bibitem[{\citenamefont{Gross}(1963)}]{JMathPhys.1963.4.195}
\bibinfo{author}{\bibfnamefont{E.~P.} \bibnamefont{Gross}},
  \bibinfo{journal}{J.Math.Phys.} \textbf{\bibinfo{volume}{4}},
  \bibinfo{pages}{195} (\bibinfo{year}{1963}).

\bibitem[{\citenamefont{Pitaevskii}(1961)}]{JETP.1961.2.451}
\bibinfo{author}{\bibfnamefont{L.~P.} \bibnamefont{Pitaevskii}},
  \bibinfo{journal}{Soviet Phys. JETP} \textbf{\bibinfo{volume}{13}},
  \bibinfo{pages}{451} (\bibinfo{year}{1961}).

\bibitem[{\citenamefont{Feynman}(1960)}]{feynman-ces60}
\bibinfo{author}{\bibfnamefont{R.~P.} \bibnamefont{Feynman}},
  \bibinfo{journal}{Caltech Engineering and Science}  (\bibinfo{year}{1960}),
  \bibinfo{note}{this is a transcript of Feynman's talk given on December 29,
  1959 at the annual meeting of the American Physical Society}.

\bibitem[{\citenamefont{Feynman}(1982)}]{feynman-82}
\bibinfo{author}{\bibfnamefont{R.~P.} \bibnamefont{Feynman}},
  \bibinfo{journal}{International Journal of Theoretical Physics}
  \textbf{\bibinfo{volume}{21}}, \bibinfo{pages}{467} (\bibinfo{year}{1982}).

\bibitem[{\citenamefont{Feynman}(1985)}]{feynman-85}
\bibinfo{author}{\bibfnamefont{R.~P.} \bibnamefont{Feynman}},
  \bibinfo{journal}{Optics News} \textbf{\bibinfo{volume}{11}},
  \bibinfo{pages}{11} (\bibinfo{year}{1985}).

\end{thebibliography}
\end{document}